\documentclass[12pt,iop]{emulateapj}

\usepackage{longtable}
\usepackage{color}
\usepackage{rotating}
\usepackage{natbib}
\usepackage[pdftitle={Constraining Radiation and Plasma Environment of KBHZ},
  colorlinks = true, breaklinks = true, citecolor = blue, linkcolor =
  blue, urlcolor = blue, pdfauthor = {BHM Gang}]{hyperref}
\newcommand{\BHMcalc}{{\tt BHMcalc }}
\newcommand{\Tau}{\mathcal{T}}
\newcommand{\Mt}{M_{\rm targ}}
\newcommand{\Lt}{L_{\rm targ}}
\newcommand{\Rt}{R_{\rm targ}}
\newcommand{\Mf}{M_{\rm field}}

\newcommand{\beq}[1]{\begin{equation}\label{#1}}
\newcommand{\eeq}{\end{equation}}

\newcommand{\Msun}{\mbox{$M_{\odot}$}}

\newcommand{\sub}[1]{_{\rm #1}}

\newcommand{\hl}[1]{#1}
\newcommand{\hls}[1]{#1}

\begin{document}

%%%%%%%%%%%%%%%%%%%%%%%%%%%%%%%%%%%%%%%%%%%%%%%%%%%%%%%%%%%%%%%%%%%%%%%%%%%%%%%%
% FRONTMATTER
%%%%%%%%%%%%%%%%%%%%%%%%%%%%%%%%%%%%%%%%%%%%%%%%%%%%%%%%%%%%%%%%%%%%%%%%%%%%%%%%

\title{Constraining the Radiation and plasma environment of the {\it Kepler}
  Circumbinary Habitable-Zone Planets}

%\title{ Kepler Circumbinary Habitable-Zone Planets}
\author{Jorge I. Zuluaga\altaffilmark{1,2}, Paul
  A. Mason\altaffilmark{3}, Pablo
  A. Cuartas-Restrepo\altaffilmark{2}}

\altaffiltext{1}{Harvard-Smithsonian Center for Astrophysics,
  Cambridge, MA 02138, USA}
\altaffiltext{2}{FACom - Instituto de F\'{\i}sica - FCEN, Universidad
  de Antioquia, Calle 70 No. 52-21, Medell\'{\i}n, Colombia}
%\altaffiltext{3}{Department of Physics, University of Texas at El
%  Paso, El Paso, TX 79968, USA}
\altaffiltext{3}{New Mexico State University - 
DACC, Las Cruces, NM, 88003, USA}

\begin{abstract}

The discovery of many planets using the {\it Kepler} telescope
includes ten planets orbiting eight binary stars. Three binaries,
Kepler-16, Kepler-47, and Kepler-453, have at least one planet in the
circumbinary habitable-zone (BHZ).  \hls{We constrain the level of
  high-energy radiation and the plasma environment in the BHZ of these
  systems}. With this aim, BHZ limits in these Kepler binaries are
calculated as a function of time, and the habitability lifetimes are
estimated for hypothetical terrestrial planets and/or moons within the
BHZ.  With the time-dependent BHZ limits established, a
self-consistent model is developed describing the evolution of stellar
activity and radiation properties as proxies for stellar aggression
towards planetary atmospheres. Modeling binary stellar rotation
evolution, including the effect of tidal interaction between stars in
binaries is key to establishing the environment around these systems.
We find that Kepler-16 and its binary analogs provide a plasma
environment favorable for the survival of atmospheres of putative
Mars-sized planets and exomoons.  Tides have modified the rotation of
the stars in Kepler-47 making its radiation environment less harsh in
comparison to the solar system. \hls{This} is a good example of the
mechanism first proposed by Mason et al. \hls{Kepler-453} has an
environment similar to that of the solar system with slightly better
than Earth radiation conditions at the inner edge of the
BHZ. \hl{These results can be reproduced and even reparametrized as
  stellar evolution and binary tidal models progress, using our online
  tool \url{http://bhmcalc.net}}.

\end{abstract} 


\maketitle

%%%%%%%%%%%%%%%%%%%%%%%%%%%%%%%%%%%%%%%%%%%%%%%%%%%%%%%%%%%%%%%%%%%%%%%%%%%%%%%%
\section{Introduction}
\label{sec:introduction}
%%%%%%%%%%%%%%%%%%%%%%%%%%%%%%%%%%%%%%%%%%%%%%%%%%%%%%%%%%%%%%%%%%%%%%%%%%%%%%%%

The outstanding work of the {\it Kepler Observatory} circumbinary team
has resulted in the discovery of \hl{10 \hls{confirmed} planets}
around eight moderately separated binaries (all have binary periods in
the range of 8-60 days).  These are Kepler-16 \citep{Doyle11},
Kepler-34 and Kepler-35 \citep{Welsh12}, Kepler-38 \citep{Orosz12b},
Kepler-47 \citep{Orosz12a}, Kepler-64 \citep{Kostov13,Schwamb13},
Kepler-413 \citep{Kostov14}, Kepler-47 \citep{Hinse14} and
\hls{Kepler-453 \citep{Welsh15}}.  Interesting, three out of eight of
\hl{these systems} ($38\%$), namely, Kepler-16, Kepler-47, and
\hls{Kepler-453}, host giant planets in the circumbinary
habitable-zone (BHZ)\footnote{\hls{Recently, the discovery of a third
    planet around Kepler-47.  The discovery of this planet may imply a
    modification of the properties of Kepler-47c as considered
    here.}}.

The study of conditions that planets face while orbiting a binary
system has attracted the attention of the exoplanetary community.
\hls{Several authors have studied} the stability of their orbits
\citep{Dvorak86, Holman99} and the kind and level of insolation
experienced by planets while illuminated by two stars
\citep{Harrington77, Haghighipour13}. \hl{The conditions for the
  formation and migration of these planets have been studied by
  \citet{Pierens08}, \citet{Kley14}, and
  \hls{\citet{Kley15}}}. Efforts are being made to constrain the
magnetic and plasma environment of circumbinary planets
\citep{Mason12, Mason13, SanzForcada14}, \hls{finding that binaries}
could potentially pose severe restrictions on the habitability of
Earth-like planets and exomoons.  \hls{We now know that} stellar
aggression, in the form of high-energy radiation and
coronal-activity-related mass-loss, would have an effect on the
evolution of planetary atmospheres, especially their capacity to
retain water (see, e.g. \citealt{Lammer13}).

In \citet{Mason13} we proposed a mechanism for which some BHZ planets
experience Earth-like or lower levels of stellar aggression and
thereby have the potential to retain water. Tidal torquing of stellar
rotation, especially for certain binary periods and initial
eccentricities, aids in the reduction of stellar activity. The
reduction in intensity of rotationally driven dynamo action results in
a decrease in stellar coronal activity, potentially promoting life on
planets.  \hls{We refer to the sum of these and other beneficial
  effects for life on planets around binaries as the {\it Binary
    Habitability Mechanism} (BHM)}.

Recent observations lend support to the tidal astrophysics of BHM. For
example, Wasp-18, a 0.63 Gyr old F6 star, is orbited by a nearby giant
planet, with a 0.94 day orbital period. The star shows a low level of
activity, normally expected for a much older star
\citep{Pillitteri14}. This effect has been described as ``premature
aging.''  In another example, the interaction of HD 189733 with a
close-in planet, has resulted in the opposite effect on a much larger
timescale \citep{Wolk11}.  \hl{HD 189733, a relatively old star,
  displays anomalously high activity levels, lending support to our
  prediction that early tidal rotational torquing could result in a
  high level of activity.  We call the latter phenomenon the ``forever
  young'' effect, in reference to the fact that tidal interaction of a
  star with a stellar or substellar companion could make the star very
  active (to appear young) for a longer time than expected from
  single-star evolution} \citep{Mason13,SanzForcada14}.

Although the specific physical mechanisms operating in the case of
Wasp-18 and HD 189733 may be different from those predicted in stellar
binary systems, these discoveries confirm the necessity of including
tidal interaction in order to assess stellar activity levels of stars
with companions and its effect on their planets.

In previous work, we used first order estimation of tidal
synchronization times \citep{Mason13}. \hls{More recently
  \citep{Mason15}}, we explored habitable niches around hypothetical
circumbinary environments, using a basic model for stellar rotation
evolution including stellar tides \hls{but} applied only to the early
phases of main-sequence evolution.

Here \hls{our} models are refined by (1) extending the time frame to
the critical pre-main-sequence phase where rotational and activity
evolution is much richer; (2) applying a more physically motivated
model for stellar rotational evolution, including but not restricted
to angular momentum (AM) transport inside the star; and (3) estimating
mass loss and X-ray emission using the latest solar-inspired models,
relating basic stellar properties and rotation to chromospheric and
coronal activity \citep{Cranmer13}.

\hls{For these three critical refinements, we are applying models that
  have been developed and tested in the case of single-stars.  We are
  assuming here that the same models can be applied to describe
  low-mass stars in moderately separated binaries (orbital periods
  larger than 8 days and separations larger than several tens of
  stellar radii). We will justify and test this assumption later on in
  the paper.}

\hls{We apply our improved and extended model to the interesting and
  well-characterized {\it Kepler} binaries with BHZ planets.}
\hl{\hls{The planets currently known in those systems} are Neptune to
  Saturn sized, so no surface habitability is expected}. \hl{However,}
it is reasonable to suspect that Earth-like planets may exist \hl{in
  the BHZ of similar binaries \citep{Eggl13}}.  Additionally, since
circumbinary giant planets appear to be common, some of them may
harbor potentially habitable exomoons \citep{Heller14}.  These planets
and putative moons are expected to experience \hl{harsh environments
  from aggressive host stars. Thus, understanding the radiation and
  plasma environment experienced by these bodies is a key effort to
  accessing their potential habitability \citep{Zuluaga12,Heller13}}.

The present aim is to constrain the radiation and plasma environment
of these HZs and to investigate the \hls{effect that these conditions
  have} on the atmospheric evolution of hypothetical \hls{terrestrial
  planets and exomoons.}

This model can readily be applied to other {\it Kepler} binaries or any
other \hls{system} with well-determined parameters. Future telescopes
like {\it PLATO} \citep{Rauer13} will likely discover circumbinary
planets exhibiting some potential for habitability.  \hls{Our aim here
  is to inform such searches by showing how a more comprehensive
  theory of planetary habitability can be applied to circumbinary
  planets especially in the presence of interacting stars.}

In order to provide the reader an opportunity to reproduce these
results and to further explore the vast parameter space of
circumbinary planets, a Web-based {\it BHM Calculator}, \BHMcalc, is
made available using the link \url{http://bhmcalc.net}.  In
\citet{Mason15} we introduced the first version of the tool.  Here we
present a comprehensive version of the calculator, detailing all the
new physical effects described in this paper, and providing an
improved and user-friendly interface.

Other online tools have recently been made available
\citep{Cuntz14,Muller14}.  Like these, the \BHMcalc provides
renditions of the instantaneous and continuous BHZ.  In addition,
\BHMcalc calculates time-dependent stellar and planetary properties
and utilizes them to study HZ environments. Most critically, the
rotational evolution of the stellar components is derived, allowing
for estimates of the resulting stellar mass loss and magnetic
activity. Stellar emission is used to evaluate effects experienced by
planets. Specifically, we select the \hls{integrated X-ray and
  extreme-ultraviolet radiation (XUV)} and \hls{stellar wind (SW)}
fluxes as derived for planets, located over a range of distances from
the binary center of mass, as proxies for habitability.  \hls{For this
  purpose we estimate} the XUV- and SW-induced planetary atmospheric
mass loss.  \hl{In all cases, the evolution of each star is calculated
  independently and orbital averages are used to calculate the
  combined effect of the binary on the planet.}

This paper is organized as follows: In Section \ref{sec:KBHZP}, the
sample of three {\it Kepler} binaries possessing BHZ
planets is described. Section \ref{sec:BHZ} \hls{presents an efficient
  model to} calculate BHZs and their evolution in time.  In Section
\ref{sec:AtmosphericEvolution}, we revisit the argument for the need
to access the radiation and plasma environment of planets as potential
drivers of atmospheric evolution. High-energy radiation and winds are
key to constraining habitability.  The comprehensive model used to
calculate the evolution of rotation and activity is described in
detail in Section \ref{sec:EvolutionRotationActivity}.  The model is
applied to constrain the radiation and plasma environments of the
solar system as a reference in Section \ref{sec:SolarSystem}. The
binary rotational evolution affects potential circumbinary worlds and
so is presented in Section \ref{sec:RotationalEvolutionKeplerStars},
and the radiation and plasma environments are investigated in Section
\ref{sec:RadiationPlasmaEnvironments}. Finally, in Section
\ref{sec:SummaryConclusions}, a summary and the conclusions of this
investigation are presented.

\label{sec:RotationalEvolutionKeplerStars}

%%%%%%%%%%%%%%%%%%%%%%%%%%%%%%%%%%%%%%%%%%%%%%%%%%%%%%%%%%%%%%%%%%%%%%%%%%%%%%%%
\section{{\it Kepler} Circumbinary Habitable-Zone Planets}
\label{sec:KBHZP}
%%%%%%%%%%%%%%%%%%%%%%%%%%%%%%%%%%%%%%%%%%%%%%%%%%%%%%%%%%%%%%%%%%%%%%%%%%%%%%%%

It is interesting to note that while none of the currently known
transiting circumbinary planets are Earth-like or even super-Earths,
this may be due purely to three selection effects. The Kepler
telescope is most sensitive to the detection of large planets,
orbiting small stars, on short-period orbits. However, the fact that
on average circumbinary planets have longer periods than those Kepler
planets with single-star hosts is not a selection effect.
Short-period circumbinary planets are limited by dynamical stability
\citep{Harrington77, Dvorak86, Holman99} and \hl{probably also because
  of formation \citep{Pierens08,Kley14}}. Remarkably, the BHZ extends
well beyond the orbital stability limit in many cases (see
\citealt{Mason15}).  In addition, residence of planets in the BHZ
appears to be common, as 3 of the 10 known transiting planets are
located in the BHZ and three binaries out of eight harbor a BHZ
planet.

In Table \ref{tab:KBHZ}, properties of the three transiting BHZ
planets and their host stars are listed.  Hereafter, we will refer to
this as the {\it KBHZ sample}.  The binaries in the {\it KBHZ sample}
are among the best-characterized binaries known, mainly owing to the
extraordinary observational constraints provided by transiting
circumbinary planets.

%TTTTTTTTTTTTTTTTTTTTTTTTTTTTTTTTTTTTTTTTTTTTTTTTTTTTTTTTTTTTTTTTTTTTTTTTTTTTTTTT
%TABLE 1
%TTTTTTTTTTTTTTTTTTTTTTTTTTTTTTTTTTTTTTTTTTTTTTTTTTTTTTTTTTTTTTTTTTTTTTTTTTTTTTTT
\begin{table*}
\begin{center}
\small
%\bigskip
\begin{tabular}{lccc}
\hline\hline
Parameter & Kepler-16b  & Kepler-47c  & Kepler-453b  \\
\hline\multicolumn{4}{c}{Stellar Properties}\\\hline
$M_{\rm 1} \ (M_{\odot})$  & 0.6897$\pm$0.0034 & 1.043$\pm$0.055 & 0.944$\pm$0.010 \\
$R_{\rm 1}\ (R_{\odot})$  & 0.6489$\pm$0.0013 & 0.964$\pm$0.017 & 0.833$\pm$0.011 \\
$T_{\rm eff_{1}}$ (K) & 4450$\pm$150 & 5636$\pm$100 & 5527$\pm$100 \\
$P_{\rm rot,1}$ (d) & 35.1$\pm$1.0 & 7.775$\pm$0.022 & 20.31$\pm$0.47\\
$M_{\rm 2} \ (M_{\odot})$  & 0.20255$\pm$0.00066 & 0.362$\pm$0.013 & 0.1951$\pm$0.0020 \\
$R_{\rm 2} \ (R_{\odot})$  & 0.22623$\pm$0.00059 & 0.350$\pm$0.006 & 0.2150$\pm$0.0014 \\
$T_{\rm eff_{2}}$ (K) & ... & 3357$\pm$100 & 3226$\pm$100 \\
$P_{\rm rot,2}$ (d) & ... & ... & ...\\
$[$Fe/H$]$ (dex) & -0.20 & -0.25 & -0.34 \\
\hline\multicolumn{4}{c}{Closest Evolutionary Track}\\\hline
$M^*_{\rm 1} \ (M_{\odot})$  & 0.70 & 1.05 & 0.90\\
$M^*_{\rm 2} \ (M_{\odot})$  &  0.20 & 0.35 & 0.20 \\
$Z^*_{\rm 1} \ (M_{\odot})$  &  0.010 & 0.010 & 0.008 \\
$\Delta \tau^*_{\rm MS,1}$ (Gyr) & $>$ 13 & 5.53 & 10.1 \\
\hline\multicolumn{4}{c}{Binary Properties}\\\hline
$T_{age}$ (Gyr) & 2.0--3.0 & 1.0--5.0 & 1.5--2.5 \\
$P_{\rm bin}$ (d)  & 41.07922$\pm$0.00007 & 7.44837695$\pm$0.00000021 & 27.322037$\pm$0.000017 \\
$a_{\rm bin}$ (AU)  & 0.22431$\pm$0.00035 & 0.0836$\pm$0.0014 & 0.18539$\pm$0.00066 \\
$e_{\rm bin}$  & 0.15944$\pm$0.00062 & 0.0234$\pm$0.0010 & 0.0510$\pm$0.0037 \\
\hline\multicolumn{4}{c}{Planetary Orbit}\\\hline
$P_{\rm p}$ (d) & 228.776$\pm$0.029 & 303.158$\pm$0.072 & 240.503$\pm$0.053 \\
$a_{\rm p}$ (AU)  & 0.7048$\pm$0,0011 & 0.989$\pm$0.016 & 0.7903$\pm$0.0028 \\
$e_{\rm p}$ & 0.0069$\pm$0.0012 & $<$ 0.411 & 0.0359$\pm$0.0088 \\
\hline\multicolumn{4}{c}{References}\\\hline
\null & \citealt{Doyle11, Winn11} & \citealt{Orosz12b} & \citealt{Welsh15} \\
\hline\hline
\end{tabular}
\caption{\small The {\it KBHZ sample}, {\it Kepler} binary systems harboring
  planets in the BHZ.  Note. Quantities
  marked with an ``*'' are model parameters.\label{tab:KBHZ}}
\end{center}
\end{table*}
%TTTTTTTTTTTTTTTTTTTTTTTTTTTTTTTTTTTTTTTTTTTTTTTTTTTTTTTTTTTTTTTTTTTTTTTTTTTTTTTT

To study the binaries in the {\it KBHZ sample}, we focus on three
different scenarios, selected to answer the following questions: (1)
If the actual circumbinary planets were Earth-like planets instead of
giant planets, then would those planets be habitable?\footnote{\hl{It
    is important to stress here that this is an hypothetical scenario.
    We are not under any circumstance assuming that the discovered
    giants planets can be habitable or that terrestrial planets also
    exist in the BHZ of all of them.}}. From necessity, this must be
based on current understanding of Earth-like habitability. (2)
Assuming that the actual circumbinary planets, which have masses
larger than or around that of Saturn, would be able to harbor a
Mars-sized exomoon, \hl{then would these moons be able to retain a
  dense and moist atmosphere in order to sustain life?} And maybe most
importantly, binaries like these are very common, so (3) what are the
constraints on habitability of currently unseen planets within the BHZ
of similar binaries?  Insight into these and other questions is gained
by studying well-observed cases, where the environmental conditions
experienced by planets under the reign of the binary may be
investigated.

It is important to stress that the current investigation is not
restricted by dynamical or formation constraints, which may or may not
limit the likelihood of these scenarios \citep{Chavez13, Chavez15,
  Andrade14, Forgan15, Pilat14}. The basic dynamical constraint
imposed
%Jaime14 by the critical limit surrounding the binary, beyond which
planets have stable orbits is included as it extends into the BHZ in
some cases (see Section \ref{sec:BHZ}).  We assume that exomoons and
multiple planets could lie in stable orbits within the BHZ of these
binaries, based on \hl{our own basic tests using numerical orbital
  integrations}.

%%%%%%%%%%%%%%%%%%%%%%%%%%%%%%%%%%%%%%%%%%%%%%%%%%%%%%%%%%%%%%%%%%%%%%%%%%%%%%%%
\section{The Circumbinary Habitable-Zone}
\label{sec:BHZ}
%%%%%%%%%%%%%%%%%%%%%%%%%%%%%%%%%%%%%%%%%%%%%%%%%%%%%%%%%%%%%%%%%%%%%%%%%%%%%%%%

Estimates of BHZ boundaries quickly followed the {\it Kepler} discoveries
\citep{Mason12, Quarles12, Haghighipour13, Kane13, Mason13}.  Our
approach, as that of others, is to use the models of
\citet{Kasting93}, updated by \citet{Kopparapu13} derived for single
stars to \hl{also} estimate the more complex BHZ limits. \hl{An
  improved version of the first-order method presented in
  \citet{Mason13} is developed and applied here}.

HZ limits are determined by the critical fluxes at which planetary
surface temperatures are compatible with the existence of liquid
water. Using one-dimensional atmospheric models, \citet{Kopparapu13}
calculated critical fluxes assuming that the planet is exposed to
black-body radiation with an effective temperature $T\sub{eff}$:

\begin{equation}
\label{eq:SHZ}
S\sub{eff,i}\equiv
S\sub{eff,i,\odot}+a\sub{i}T_{*}+b\sub{i}T^{2}_{*}+c\sub{i}T^{3}_{*}+d\sub{i}T^{4}_{*}
\end{equation}

Here $T_{*}=T_{\rm eff}-5780 K$, and $S\sub{\rm eff,i}$ and
$a\sub{i},b\sub{i},c\sub{i},d\sub{i}$ are the effective critical flux
for a solar-mass star and the interpolation coefficients for the $i$th
boundary (see Table 3 in \citealt{Kopparapu13E}).

The limits of the HZ, around either single-stars or a binary, are
calculated by finding the radius of a circle centered in the
barycenter of the system, where the average flux $\langle S \rangle$
received from the star or the binary is equal to the critical flux
calculated with Eq. (\ref{eq:SHZ}).

In the case of a single-star, $\langle S \rangle$ is equal to the
nearly constant flux $S\sub{sing}=L/d^2$ measured at distance $d$.
Here $L$ is the bolometric luminosity of the star measured in solar
units and $d$ is in astronomical units (AU).  \hls{More precisely and
as we will explain later, $L$ is a corrected luminosity that takes
into account the response of the atmosphere to the incident stellar radiation
(see Eq. \ref{eq:Wfactor})}.

The radius of the $i$th HZ edge $l\sub{sing,i}$ will then be given by:

\beq{eq:lHZ}
S\sub{eff,i}=\frac{L}{l\sub{sing,i}^2}
\eeq

Around binaries, planetary atmospheres, even if \hls{planets are} in a
circular orbit, will be subject to a variable flux and spectrum of
radiation.  The instantaneous bolometric flux on a circumbinary planet
is calculated as

\beq{eq:BinaryFlux}
S\sub{bin}(d,\theta)=\frac{L_1}{R_1^2(d,\theta)}+\frac{L_2}{R_2^2(d,\theta)}
\eeq

\noindent where $R_1$ and $R_2$ are the distances from the stars to a
point in a circle of radius $d$.  $\theta$ is the angle between the
line joining the stars and a fixed point on the circle.  This angle is
a function of time, and it will depend on the relative angular
velocity of the binary and a test particle on the circle.

$R_1$ and $R_2$ are given by

\begin{eqnarray*}
R_1^2(d,\theta) & = & (d-r_2\sin\theta)^2+r_2^2\cos^2\theta\\
R_2^2(d,\theta) & = & (d+r_1\sin\theta)^2+r_1^2\cos^2\theta
\label{eq:R1R2}
\end{eqnarray*}

\noindent where $r_1=q\,r/(1+q)$ and $r_2=q\,r_1$ are the
instantaneous distances of the stars to the barycenter and $q = M_2
/M_1$ is the binary mass ratio.  The relative distance $r$ between the
stars is also a function of time for the general case of an eccentric
binary orbit.

Since the two stars have different effective temperatures,
$S\sub{bin}$ and $S\sub{eff,i}$ cannot be directly compared as in the
case of single-stars.  However, in the case of stellar twins, the two
sources will have the same effective temperature and their fluxes can
be summed directly.  On the other hand, in disparate binaries the
ratio of the flux of the secondary to that of the primary scales with
the fourth power of the ratio of their effective temperatures.  As a
result the combined spectrum, to first order, may be assumed to have
an effective temperature equal to that of the primary and the total
flux equals $S\sub{bin}$ \citep{Mason13}.  A verification of this
approximation is discussed below.

A more consistent method to take into account the differences in
effective temperatures of the stellar components is to replace
bolometric luminosities $L_1$ and $L_2$ in Eq. (\ref{eq:BinaryFlux})
by weighted luminosities \citep{Haghighipour13,Cuntz14}:

\beq{eq:Wfactor}
L'_{1,2}=L_{1,2} (1+[S\sub{eff,i}(T_{1,2})-S\sub{eff,i,\odot}] l\sub{i,\odot}^2)
\eeq

The weights in parenthesis account for the different spectral
distribution of each source and the response of planetary atmosphere
to these spectra.

To obtain the limits of the BHZ, $S\sub{bin}$ given by
Eq. (\ref{eq:BinaryFlux}), \hls{the equation} should be averaged over
a time long \hls{enough} compared to the periods of the system. If the
binary eccentricity is not too large, an analytical expression for
$\langle S\rangle$ may be used:

\beq{eq:Sint}
{\scriptstyle
\langle S\sub{bin}\rangle(d) = \sum_{n\neq m}
\frac{L_n'}{\sqrt{A_m^2-B_m^2}}
\left[1-\frac{1}{\pi}\arctan\left(\frac{B_m}{\sqrt{A_m^2-B_m^2}}\right)\right]
}
\eeq

\noindent with $n,m=1,2$, $A_m\equiv d^2+r_m^2$ and $B_m\equiv
2\;d\;r_m$.

Finally, the BHZ limits are calculated by solving the equation

\beq{eq:lbin}
S\sub{eff,i} =
\langle S\sub{bin}\rangle(l\sub{bin,i})
\eeq

Although other authors have developed alternative methods to estimate
the limits of the BHZ \citep{Haghighipour13, Kane13, Mason13, Cuntz14,
  Muller14}, all of them share the most important characteristics of
the method presented here.  In particular, our method is an improved
version of the first order method used in \citet{Mason13} by
incorporating some key elements of the methods presented by
\citet{Haghighipour13} and \citet{Cuntz14}.  An interesting advantage
of our method with respect to others is the fact that for low
eccentricities, the calculation of the BHZ limits requires finding the
roots of an analytical formula.  No numerical integrations or
calculations over a grid of points around the system are required.
This feature greatly improves the speed of calculation, which is
especially important for the exploration of the vast parameter space
of circumbinary habitability.

\begin{figure}
\begin{center}
\includegraphics[width=80mm]{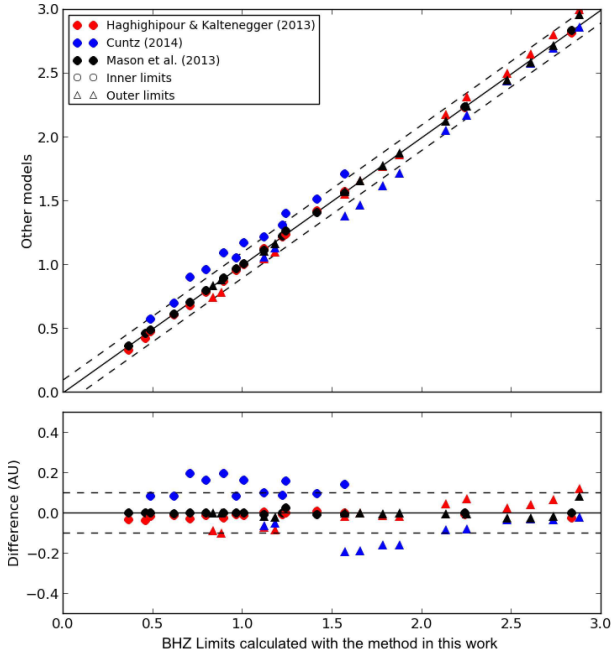}\\\vspace{0.5cm}%DONE
\caption{Comparison of BHZ limits for all the {\it Kepler} binaries having
  circumbinary planets, calculated with four different methods:
  \citet{Haghighipour13,Mason13,Cuntz14} and the current paper.  The
  limits calculated with the method presented here are used as
  reference values (abcsisa values and solid black lines in upper
  panel). Dashed lines correspond to a difference of 0.1 AU
  ($\sim$10\%).
 \label{fig:BHZ-Comparison}}
\end{center}
\end{figure}

We have compared the BHZ limits for all the {\it Kepler} binaries with known
circumbinary planets calculated using our method and those used in
three different investigations. \hls{The results are presented in
  Figure \ref{fig:BHZ-Comparison}}. We have found that the average of
the relative difference between the BHZ limits using the new method
and that of \citet{Haghighipour13} is less than 4\% \footnote{The
  method devised by these authors produces dynamical and asymmetric
  BHZ.  Comparison with our method depends on the assumptions made
  about how to convert their asymmetrical limits into our symmetric
  ones.  Part of the discrepancy may involve this ambiguity.} The
difference found by comparing our method with that of \citet{Mason13}
is less than 1\%.  The limits calculated by \citet{Cuntz14} are
systematically more conservative than all of the other methods, with
the inner limits $\sim$10\% farther out and the outer limits
$\sim$10\% closer in than the other methods.

%FFFFFFFFFFFFFFFFFFFFFFFFFFFFFFFFFFFFFFFFFFFFFFFFFFFFFFFFFFFFFFFFFFFFFFFFFFFFFFFF
%FIGURE 0
%FFFFFFFFFFFFFFFFFFFFFFFFFFFFFFFFFFFFFFFFFFFFFFFFFFFFFFFFFFFFFFFFFFFFFFFFFFFFFFFF
%FFFFFFFFFFFFFFFFFFFFFFFFFFFFFFFFFFFFFFFFFFFFFFFFFFFFFFFFFFFFFFFFFFFFFFFFFFFFFFFF

%======================================================================
\subsection{Evolution of Insolation and BHZ}
\label{subsec:evolutionBHZ}
%======================================================================

\hl{Generally, it is possible to assume that moderately separated
  stars in binary are formed at the same time.  Moreover, stellar
  evolution shows that stars with different masses evolve at different
  rates and in different ways. Thus, in order to investigate
  habitability around a binary, stellar evolution models for each star
  must be used to determine the time dependence of the BHZ.}

\hls{It is important to ask whether the fundamental properties of
  low-mass stars, such as effective temperature and luminosity used to
  calculate the BHZ, evolve in binaries exactly as they do in
  single-stars.  For moderately separated binaries, (periods larger
  than 8 days, stellar separations larger than tens of stellar radii),
  such as those studied in our sample, the Roche lobes of the stars
  are one order of magnitude larger than the stellar diameters.  For
  instance, in the case of the tightest binary in our sample
  (Kepler-47), the filling factor of the primary (ratio of stellar
  radius to Roche lobe size) is $\sim$0.2 during PMS (when the star is
  largest) and $\sim$0.1 during main-sequence \citep{Eggleton83}.  At
  those filling factors the shape of the star is only slightly
  modified by the companion and its interior and atmospheric
  properties are practically the same as for a single-star.}

\hl{HZ evolution is also a natural outcome of the evolution of
  single-stars.  In the same way that the continuous HZ is
  defined for single-stars \citep{Kasting93}, it is generalized to
  binaries and a continuous circumbinary habitable zone (CBHZ) is
  defined.}

\begin{figure}
\begin{center}
\includegraphics[width=80mm]{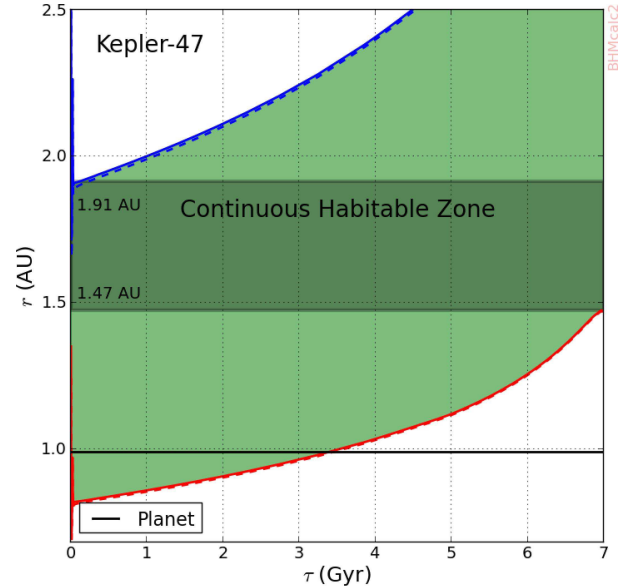}\\\vspace{0.5cm}%DONE
\caption{Kepler-47 is shown as an example of the evolution of the BHZ
  (green area).  BHZ limits are shown as a function of age (solid blue
  and red lines). The HZ for a single-star with a mass equal to the
  primary is also shown (dashed lines). Since the system is composed
  of a solar-like star and an M dwarf the limits for the single
  primary and the binary are almost the same.  The CBHZ is highlighted
  (shaded gray region; see text for explanation). The semimajor axis
  of Kepler-47c, is shown as a horizontal line. Neglecting
  eccentricity effects, the habitability lifetime of this orbit is
  roughly 3.5 Gyr.
 \label{fig:CBHZ}}
\end{center}
\end{figure}

In Figure \ref{fig:CBHZ}, \hl{the evolution of the BHZ for Kepler-47
  and its associated CBHZ is shown.  We use stellar evolutionary
  models, at each age, to calculate the instantaneous properties of
  both stars. Then the limits of the HZ at that time are determined.
  Shown here are the the Recent Venus and Early Mars criteria.}
\hls{The outer CBHZ limit is defined by the instantaneous outer edge
  of the BHZ as calculated at the time when the binary flux is
  minimum. We have tested that this time is close to the time of
  zero-age main-sequence (ZAMS) of the primary star, which by
  definition is the more massive component of the system}. Planets in
the CBHZ remain inside the BHZ during the entire main-sequence stage
of the primary as it is the star with the shortest lifetime.  The
definition of the CBHZ inner edge is trickier.  We define it as the
position of the inner edge of the instantaneous BHZ calculated when
the primary star abandons the main-sequence (close to the end of the
nuclear hydrogen-burning phase).

%FFFFFFFFFFFFFFFFFFFFFFFFFFFFFFFFFFFFFFFFFFFFFFFFFFFFFFFFFFFFFFFFFFFFFFFFFFFFFFFF
%FIGURE 1
%FFFFFFFFFFFFFFFFFFFFFFFFFFFFFFFFFFFFFFFFFFFFFFFFFFFFFFFFFFFFFFFFFFFFFFFFFFFFFFFF
%FFFFFFFFFFFFFFFFFFFFFFFFFFFFFFFFFFFFFFFFFFFFFFFFFFFFFFFFFFFFFFFFFFFFFFFFFFFFFFFF

\begin{figure*}
\begin{center}
\includegraphics[width=70mm]{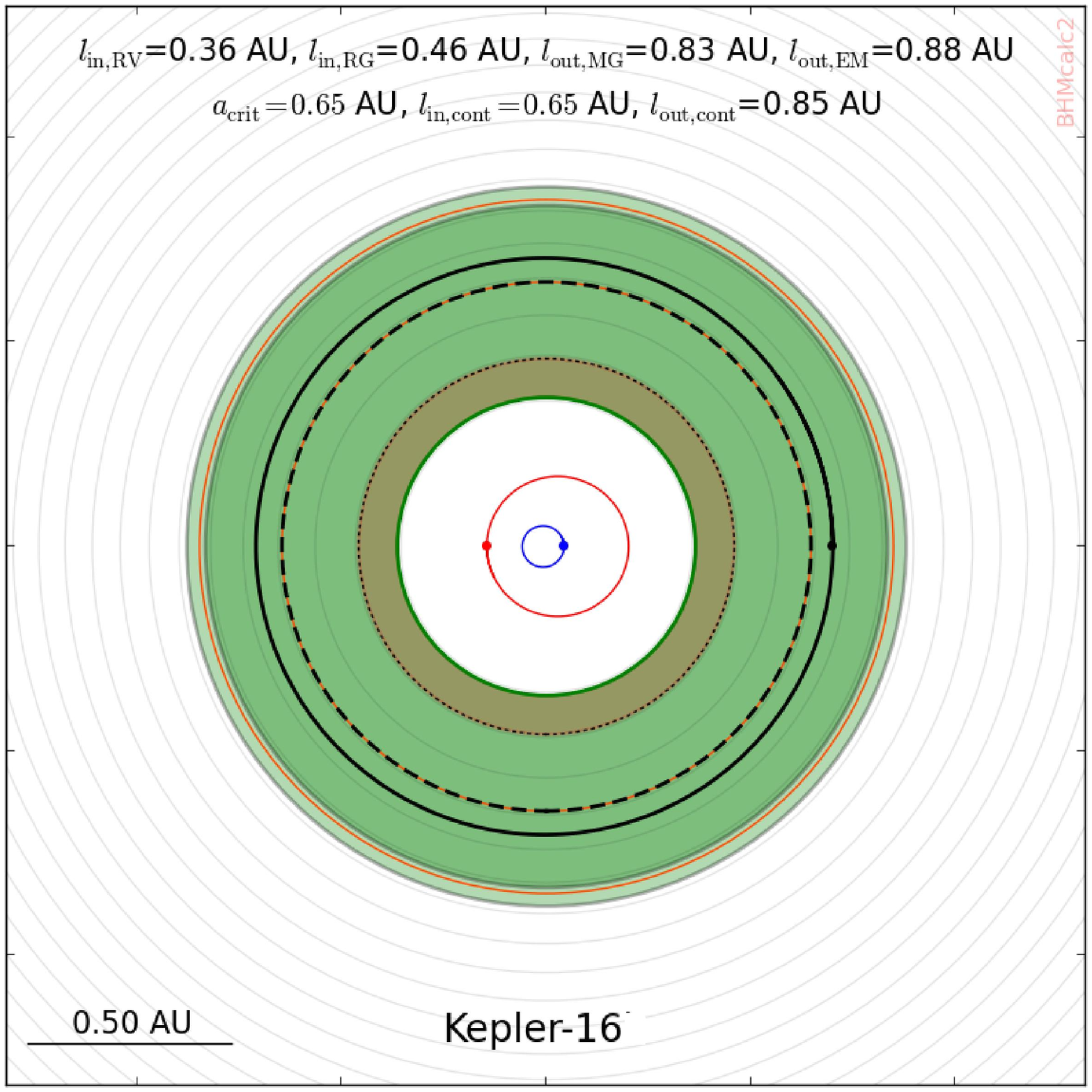}\hspace{0.5cm} %DONE
\includegraphics[width=70mm]{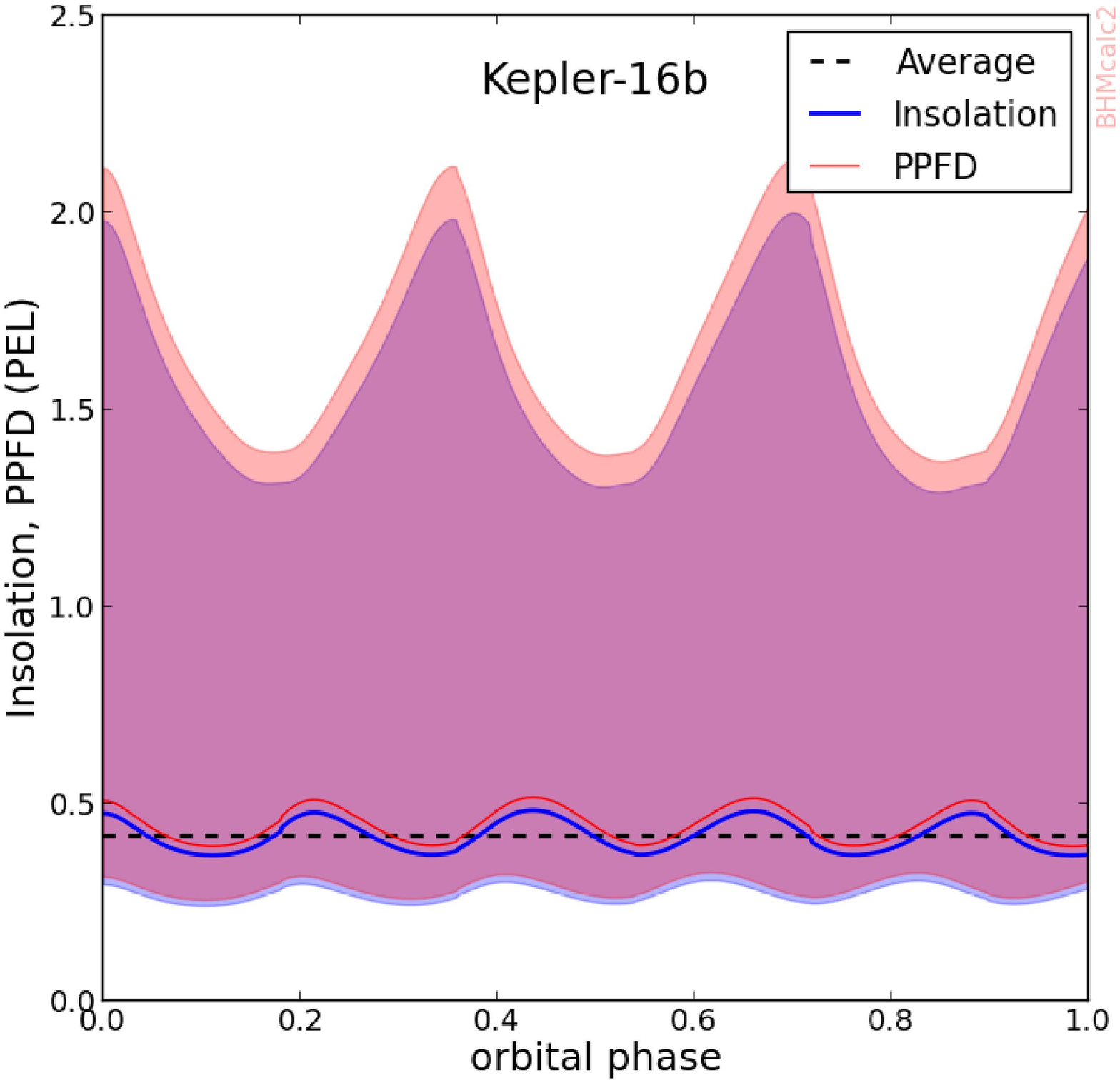}\\\vspace{0.5cm} %DONE
\includegraphics[width=70mm]{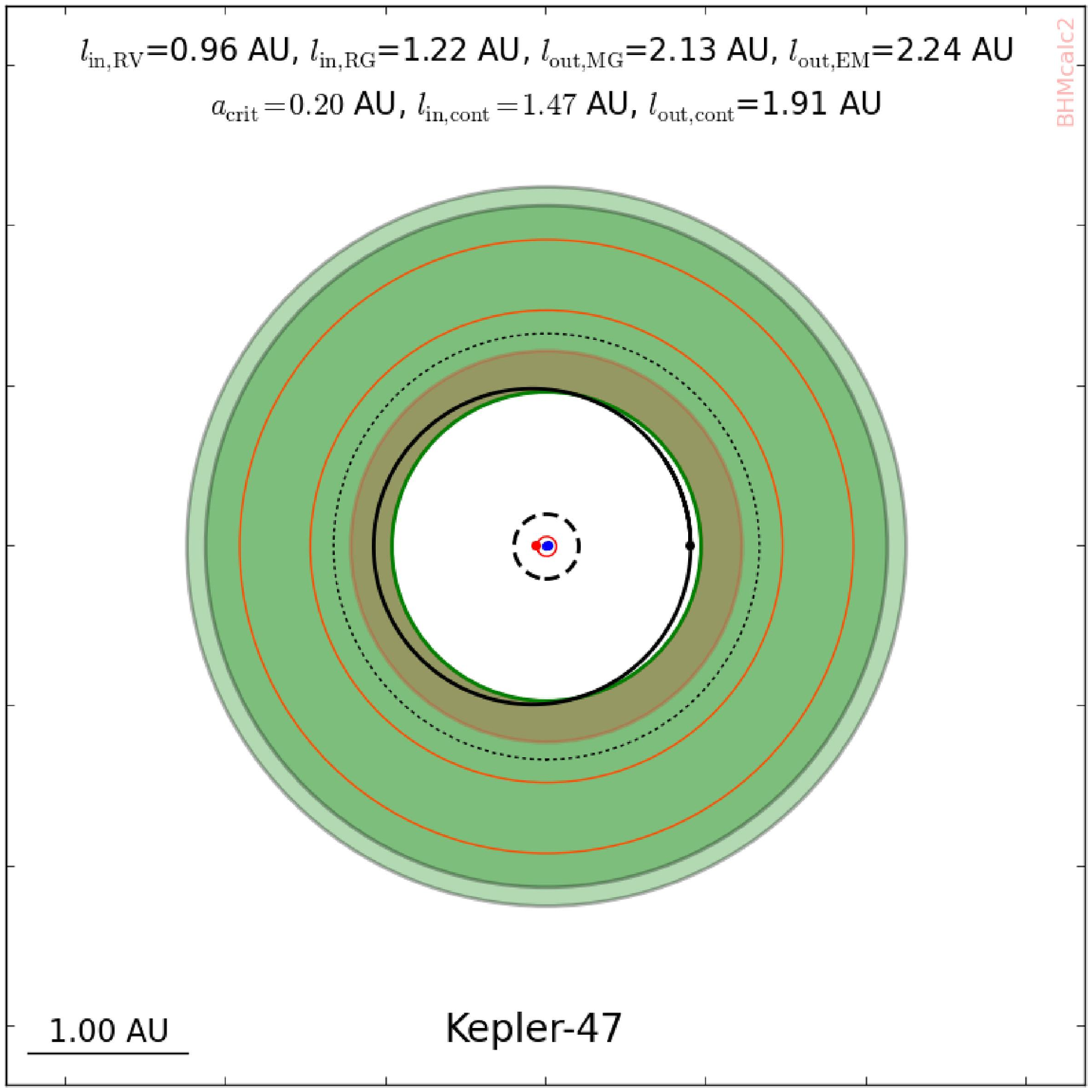}\hspace{0.5cm} % DONE
\includegraphics[width=70mm]{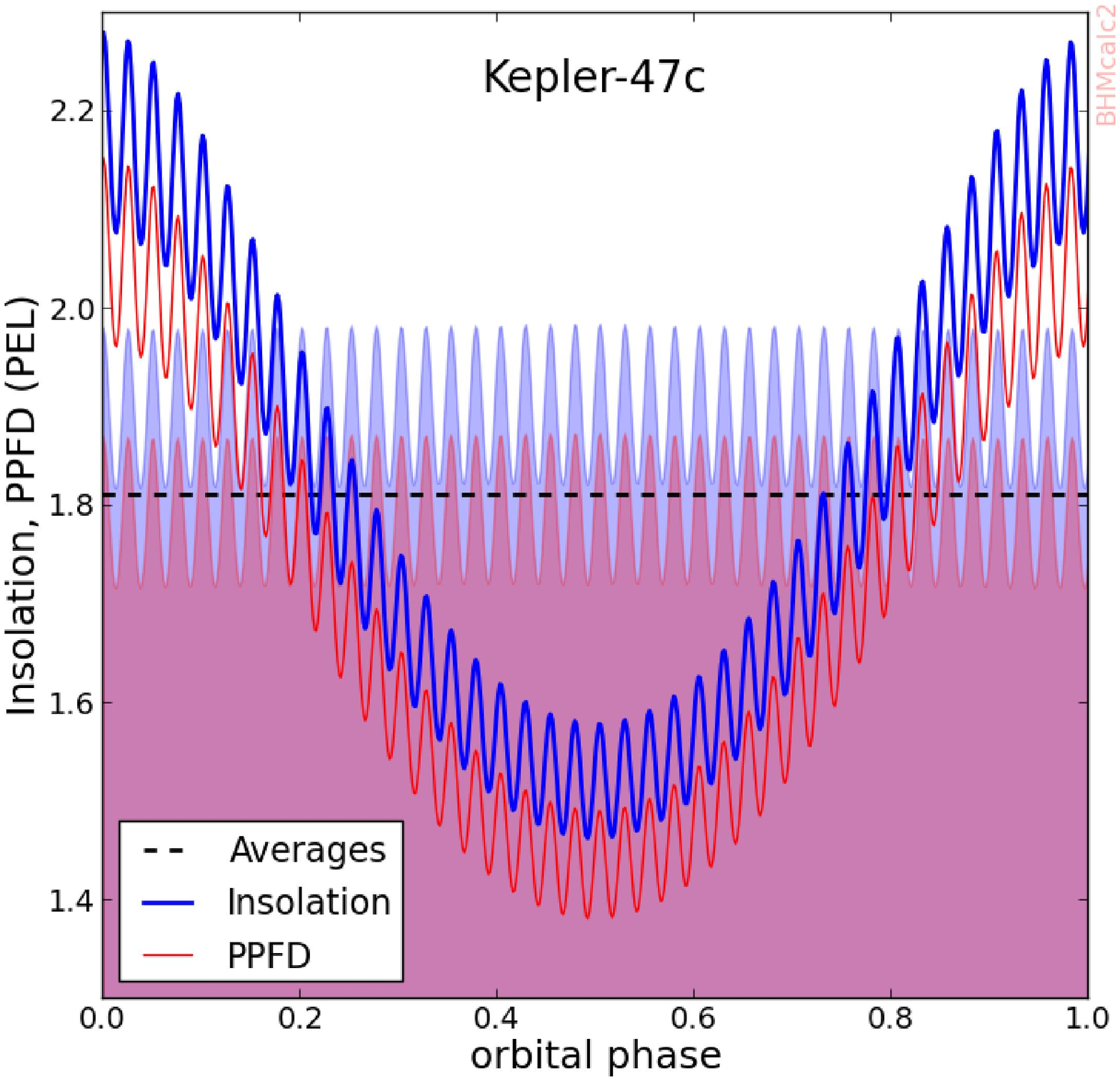}\\\vspace{0.5cm} % DONE
\includegraphics[width=70mm]{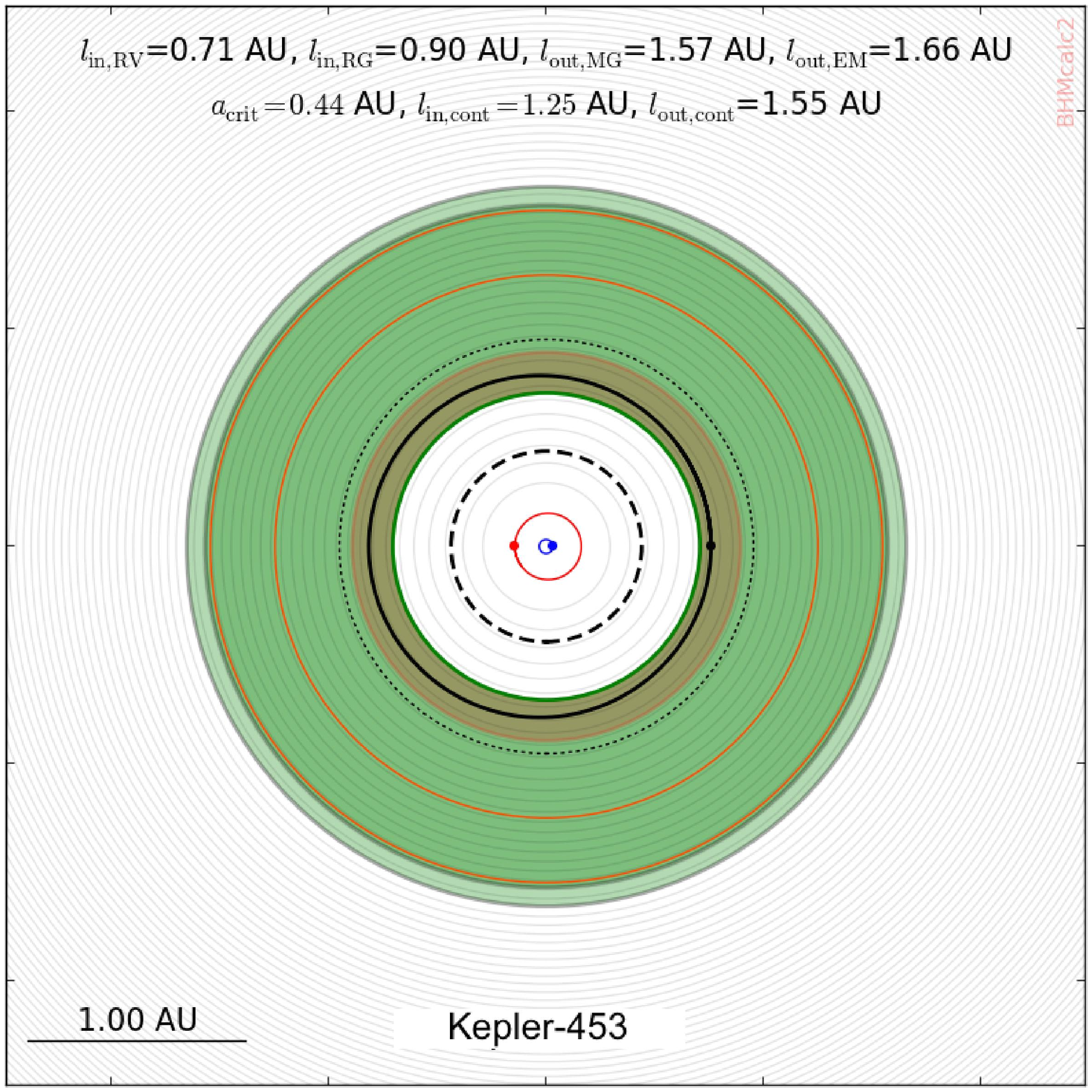}\hspace{0.5cm} %DONE
\includegraphics[width=70mm]{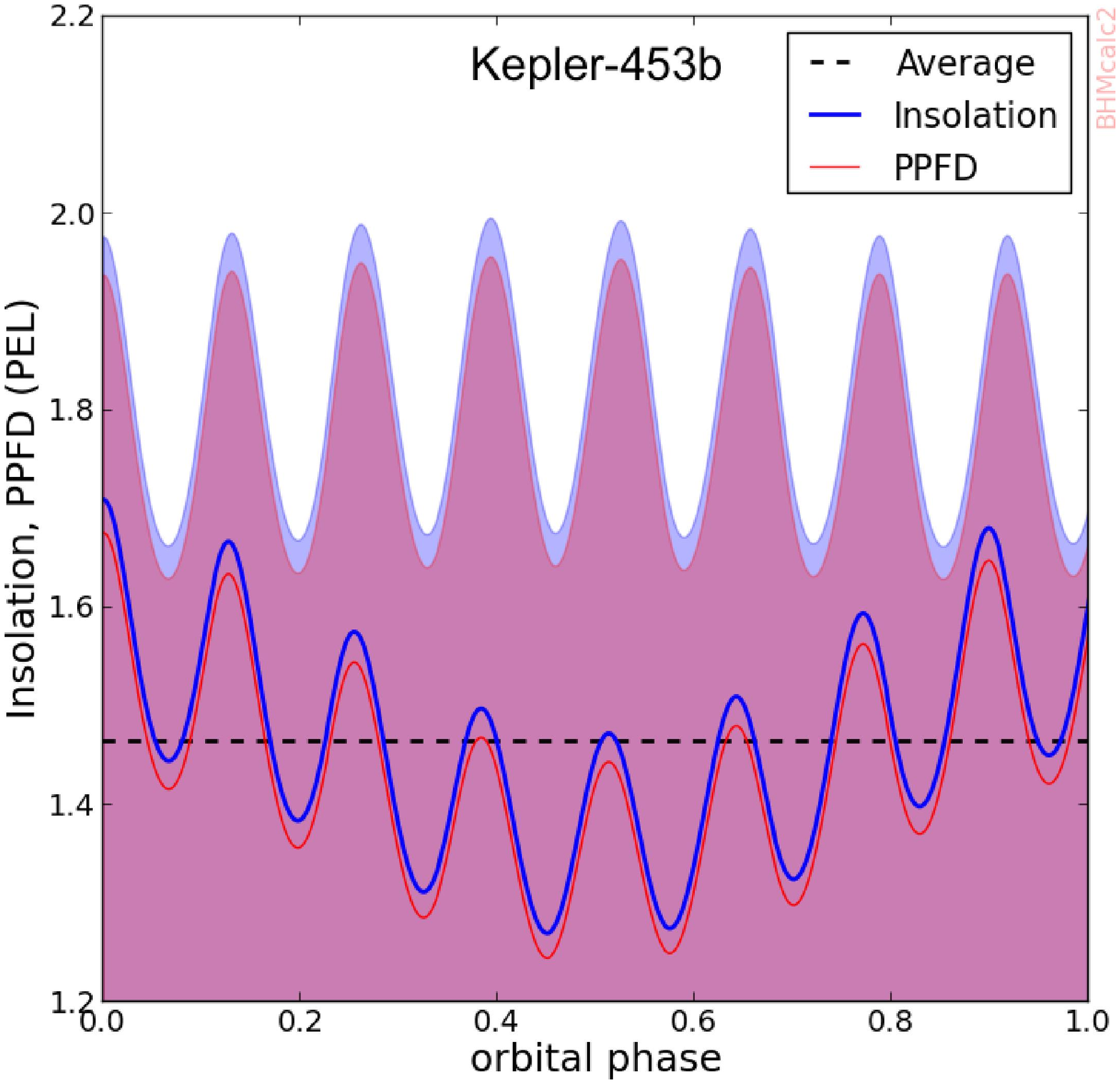}\\\vspace{0.5cm} %DONE
\caption{BHZs, insolation and PPFD (see \citet{Mason15} for an
  explanation) for the {\it KBHZ sample}. {\bf Left:} Along with the
  BHZ instantaneous limits, the binary orbit, the critical distance
  (dashed black line), the planet orbit (solid black line), the limits
  of the continuous BHZ (orange solid lines), and distances of binary
  orbit resonances (concentric gray circles) are shown. The distance
  at which the equivalent Earth insolation is received is marked with
  a dotted black line. {\bf Right:} Insolation and PPFD are calculated
  in units of present Earth levels not only at the distance of the
  planet (blue and red solid lines, respectively) but also at all
  locations within the BHZ (shaded areas).\label{fig:BHZ}}
\end{center}
\end{figure*}

In order to assess the present insolation conditions in the {\it KBHZ
  sample} we plot, in Figure \ref{fig:BHZ}, the instantaneous BHZ
calculated in the middle of the range of estimated age (See Table
\ref{tab:KBHZ} for ages and references).  Along with BHZ limits, the
insolation and photosynthetic photon flux density (PPFD) are shown as
a function of planetary orbital phase.  See \citet{Mason15} for
details on PPFD calculations.

%FFFFFFFFFFFFFFFFFFFFFFFFFFFFFFFFFFFFFFFFFFFFFFFFFFFFFFFFFFFFFFFFFFFFFFFFFFFFFFFF
%FIGURE 2
%FFFFFFFFFFFFFFFFFFFFFFFFFFFFFFFFFFFFFFFFFFFFFFFFFFFFFFFFFFFFFFFFFFFFFFFFFFFFFFFF
%FFFFFFFFFFFFFFFFFFFFFFFFFFFFFFFFFFFFFFFFFFFFFFFFFFFFFFFFFFFFFFFFFFFFFFFFFFFFFFFF

In the BHZ diagrams of Figure \ref{fig:BHZ}, the planetary orbit has
been placed among different landmark distances: BHZ and CBHZ edges,
binary orbit, equivalent Earth insolation distance, and last but not
least, the critical stability distance $a_{\rm crit}$, 
defined as the minimum distance where test particles have stable
circumbinary orbits. We compute this critical distance with the
semi-empirical relationship of \citet{Holman99}:

\begin{eqnarray}
a_{crit} & \approx & (1.60 + 5.10 e_{\rm bin} - 2.22 e_{\rm bin}^{2}+\\\nonumber
      &         & 4.12\mu - 4.27 e_{\rm bin} \mu -5.09 \mu^{2}+\\\nonumber
      &         & 4.61 e_{\rm bin}^{2}\mu^{2})\, a_{\rm bin}.
\label{eq:orbit}
\end{eqnarray}

Here $\mu=M_{2}/(M_{1}+M_{2})$, and $a_{\rm bin}$ and $e_{\rm bin}$
are the binary semimajor axis and eccentricity, respectively.
Although dynamical instability likely occurs outside the critical
limit as well, as also discussed by \citet{Holman99}, all the planets
in our sample are safely beyond this critical distance.  However, in
Kepler-16, $a_{\rm crit}$ lies inside the BHZ, restricting the
instantaneous extension of the BHZ as well as the CBHZ; see top left
panel of of Figure \ref{fig:BHZ}.  When $a_{\rm crit}$ is larger than
the inner edge of the continuous BHZ we use its value as the inner
edge distance.

In the instantaneous BHZ diagrams of Figure \ref{fig:BHZ}, distances
at which orbital resonances with the binary occur are also shown.  It
is very interesting to notice, and likely not coincidental, that in
both cases, Kepler-16b and \hls{Kepler-453}b, the planet remains
entirely in between two resonances throughout its orbit.  This
supports the idea that orbital binary resonances may play a role in
the formation and final architecture of circumbinary systems.  Notice
also that Kepler-47c crosses many tightly spaced resonances, not
plotted in its BHZ diagram in Figure \ref{fig:BHZ} for clarity.

The planet Kepler-16b has a nearly circular orbit just inside the
maximum greenhouse limit. Kepler-47c has a highly eccentric orbit and
leaves the BHZ for part of its orbit.  It does, however, currently
spend most of its time within the HZ, especially given the slower
velocity at apastron (see insolation diagram in the middle right panel
of Figure \ref{fig:BHZ}). The planet \hls{Kepler-453}b has a slightly
eccentric orbit and always remains in the HZ, just inside the runaway
greenhouse limit.

According to our calculations, of all the circumbinary {\it Kepler}
planets, only Kepler-16b is in the CBHZ, i.e. it will lie inside the
BHZ during the whole lifetime of the primary star, which is in turn
larger than the age of the universe.  This condition would seem at
first very favorable for the emergence and evolution of complex
life. However other factors could limit the long-term habitability
within environments surrounding Kepler-16 and similar binaries (see
Section \ref{sec:RadiationPlasmaEnvironments}).

Kepler-47c and \hls{Kepler-453b} do not remain inside the BHZ for the
main-sequence lifetime of the primaries. However, both remain inside
(although very close to the inner edge of) the BHZ for
astrobiologically relevant times (3.5 Gyr and 4 Gyr respectively). If
we believe stellar evolution models, Kepler-47c is about to leave or
has just left the most optimistic BHZ, being currently in a Venus-like
state in terms of stellar insolation.

In order to calculate BHZ evolution and other key quantities for
models, the stellar properties available in the {\tt PARSEC v1.2}
evolutionary tracks are used \citep{Bressan12,Bressan13}
\footnote{\href{http://people.sissa.it/\~bressan/parsec.html}{http://people.sissa.it/~bressan/parsec.html}}.
In addition to providing good coverage in metallicity and mass, the
{\tt PARSEC} tracks sample the PMS stages of stellar evolution well
\hls{(and of course also the main sequence)}, a key feature when
attempting to study the early evolution of stellar rotation.
Moreover, these models reflect improvements for low-mass stars
\citep{Chen14}.  In the {\tt BHMcalc} tool we provide access to other
evolutionary model results.

%%%%%%%%%%%%%%%%%%%%%%%%%%%%%%%%%%%%%%%%%%%%%%%%%%%%%%%%%%%%%%%%%%%%%%%%%%%%%%%%
\section{The evolution of atmospheres as a key factor for assessing habitability}
\label{sec:AtmosphericEvolution}
%%%%%%%%%%%%%%%%%%%%%%%%%%%%%%%%%%%%%%%%%%%%%%%%%%%%%%%%%%%%%%%%%%%%%%%%%%%%%%%%

So far we have calculated insolation as a way to assess habitability
conditions in circumbinary planets. But insolation is not the only
factor constraining the capability of a planet to sustain life (for a
review of many other factors see \citealt{Kasting09} and references
therein).  Among other endogenous and probably exogenous factors, the
presence of a properly dense atmosphere with the right amount of water
is mandatory.

The solar system is a good example of a planetary system where among
three planets \hl{(and a sizeable moon)}, located close or inside the
HZ, two of them, Venus and Mars \hl{(plus the Moon)}, are
un-inhabitable mainly due to factors related to the composition and
mass of their atmospheres.  

%These limiting factors likely arose during
%the first hundreds of Myrs to the first few Gyrs, after planetary
%formation.

The evolution of a planetary atmosphere is driven by the influx of
high-energy radiation and plasma from the host star. This has been a
matter of research in the last two decades (\citealt{Kasting96};
\citealt[p. 127]{Lammer07}; \citealt[p. 399]{Lammer09d};
\citealt{Lammer10}; \citealt{Lammer13}; \citealt[p.567]{Tian13}).
Although it is not yet clear which factors determine the survival of
Earth-like atmospheres or drive their composition to favor habitable
conditions, multiple lines of evidence show that the interaction of a
planet with its young host star is a strong factor in determining the
fate of potentially habitable planet atmospheres
(\citealt[p. 127]{Lammer07}; \citealt[p. 399]{Lammer09d};
\citealt{Tian09}; \citealt{Zendejas10}; \citealt{Zuluaga13}).

Two key factors are involved in this interaction. The first is the
level of X-rays and extreme-ultraviolet (EUV) radiation, collectively
called XUV radiation, incident on the planet.  These photons heat up
the upper atmosphere and under certain conditions could lead to
massive loss of volatiles (\citealt{Tian09};
\citealt[p. 567]{Tian13}).  The flux of charged particles, mainly
protons, and their associated magnetic fields (SW) is the other
factor.  The direct interaction between this continuous flow of
particles and fields and planetary atmospheres strongly impacts
unmagnetized planets \hl{and moons} (e.g., Venus, Mars \hl{and the
  moon}). Resulting nonthermal processes may remove large fractions of
planetary atmospheres, or in extreme cases \hls{they} can strip them
almost completely \citep{Zendejas10}.  Mars is the best known example
of the effect of this kind of aggression eliminating habitability.
Recent data from the {\it MAVEN} Mission
\hls{\citep{Rahmati14,Jakosky15}} are showing the increased level of
importance that SW interaction with the atmosphere has in shaping the
evolution of Mars's \hls{capability to have liquid water on its
  surface (habitability)}.

Modeling the interaction between the radiation and plasma fields of
the star and, in this case, the binary system is key in assessing the
habitability of all planets.  However, since binaries are relatively
novel environments, at least for planetary systems, we need to
understand and learn to model aspects of stellar evolution that have
been applied for years to single-stars.  In the following sections we
develop a theoretical framework to model the evolution of stellar
activity and its role in the changing conditions faced by the
atmosphere of a circumbinary habitable planet.

%%%%%%%%%%%%%%%%%%%%%%%%%%%%%%%%%%%%%%%%%%%%%%%%%%%%%%%%%%%%%%%%%%%%%%%%%%%%%%%%
\section{Stellar rotation and magnetic activity in binaries}
\label{sec:EvolutionRotationActivity}
%%%%%%%%%%%%%%%%%%%%%%%%%%%%%%%%%%%%%%%%%%%%%%%%%%%%%%%%%%%%%%%%%%%%%%%%%%%%%%%%

%%%%%%%%%%%%%%%%%%%%%%%%%%%%%%%%%%%%%%%%%%%%%%%%%%%%%%%%%%%%%%%%%%%%%%%%%%%%%%%%
\subsection{Evolution of Stellar Rotation}
\label{subsec:EvolutionRotation}
%%%%%%%%%%%%%%%%%%%%%%%%%%%%%%%%%%%%%%%%%%%%%%%%%%%%%%%%%%%%%%%%%%%%%%%%%%%%%%%%

The evolution of stellar rotation has been a matter of intense
research over the last several decades, not only in single but also in
binary stars (for a recent review see \citealt{Bouvier13} and
references there in; for the evolution of rotation in binaries see
\citealt{Claret05}).

The large diversity of stellar rotation periods observed in stars of
all ages (especially PMS stars) has driven the development of a
diverse array of semi-empirical and phenomenological models describing
the evolution of stellar AM \citep{Kawaler88,Chaboyer95,Matt12}.
These models have been extensively tested against samples of stars
(both single and binary) in young stellar clusters and more recently
in low-mass field stars \citep{Irwin11}.  The details of these models
depend on weakly constrained parameters; however, a general consensus
has been reached about the general forces and timescales driving
stellar rotation evolution.

Here we apply and extend some of these models for stars in moderately
separated binaries \hl{(i.e. binaries with similar masses and orbits
  to those of the KBHZ binaries)}.  For details concerning the
physical motivations, history of development, and success of these
models at reproducing observations, please refer to \citet{Bouvier13}
and references therein.

Although the applicability of these models to stars in binary systems
has not been systematically tested, we assume that some of the
underlying mechanisms will operate in the stellar components of these
binaries as if they were isolated stars. \hl{A detailed test of this
  hypothesis should be pursued in future research efforts.}
\hls{Still, for the range of orbital separations and stellar masses
  considered here, the evolution in insolation hypothesis remain as a
  good one for the same arguments given before to support the
  applicability of single-star evolutionary models.}

In general, the AM of the $i$th layer of a star evolves while obeying
Newton's angular second law:

\beq{eq:dWdt}
\frac{dJ_i}{dt}=\sum \Tau_i.
\eeq

Here $J_i=I_i \Omega_i$ is the instantaneous AM of the layer, $I_i$
its moment of inertia, $\Omega_i$ its instantaneous angular velocity,
and $\sum \Tau_i$ is the net torque over the layer.

In the case of low-mass stars ($M_\star<1.3 M_\odot$), it is customary
to assume that the star has two distinct interior layers: a radiative
inner core ($i={\rm core}$) and a convective envelope ($i={\rm
  conv}$).  It is also assumed that both layers rotate independently
as rigid bodies (i.e. no differential rotation).

The convective envelope is subject to two different exogenic effects:
(1) tidal \hl{breaking}, $\Tau_{\rm tid}$, and (2) loss of AM via a
magnetized SW $\Tau_{\rm SW}$.

To calculate $\Tau_{\rm tid}$, we use the formalism developed by
\citet{Hut81}.  Accordingly, a target star of mass $\Mt$ and radius
$\Rt$, rotating at an instantaneous rate $\Omega$ and affected by a
secondary star of mass $\Mf$ in an orbit with semimajor axis $a$,
eccentricity $e$, and mean angular velocity $n$, feels an effective
tidal torque given by

\begin{eqnarray}
\Tau_{\rm tid} & = & -\frac{3 k_2 I}{\kappa^2 t_{\rm
    diss}}\left(\frac{\Mf}{\Mt}\right)^2\left(\frac{\Rt}{a}\right)^6\times\\\nonumber
& & \times \frac{n}{(1-e^2)^6}\left[f_2(e^2)-(1-e^2)^{3/2} f_5(e^2)
  \frac{\Omega}{n}\right].
\label{eq:TauTidal}
\end{eqnarray}

Here $k_2$ is the apsidal motion constant that quantifies the response
of the star to tidal distortion, and it depends in general on the
internal mass distribution of the star; $t_{\rm diss}=f_{\rm
  diss}(\Mt\Rt^2/\Lt)^{1/3}$ is the timescale for dissipation of tidal
energy inside the target star, and $\kappa\equiv \sqrt{I/(\Mt\Rt^2)}$
is the gyration radius.  Eccentricity functions $f_2(e^2)$ and
$f_5(e^2)$ are defined by Eqs. (2) and (3) in \citet{Hut81}.

We assume that while tides are raised over the whole star, only tides
within the convective layer contribute to AM dissipation
\citep{Zahn08}.  As a result, the torque in Eq. \ref{eq:TauTidal} only
affects the envelope of the star.  Alternatively, we can assume that
tides raised in the radiative core are dissipated on timescales much
larger than that in the convective zone, rendering the resulting
torque comparatively small.

The parameter $f_{\rm diss}$ quantifies the efficiency of tidal energy
dissipation.  Although a fundamental theory of this process capable of
correctly explaining observations of close binaries is still lacking
(see \citealt{Claret11} for a discussion) we assume for simplicity
that $f_{\rm diss}\approx 1$, as is used, for example, in
\citet{Claret12}.  Alternative theories, predict values of $f_{\rm
  diss}$ within one order of magnitude.  Thus, for instance, the
turbulent dissipation theory of \citet{Zahn08} predicts a value
$f_{\rm diss}\approx 3.5$.

For nonnegligible eccentricities, tidal braking could drive stars to a
pseudosynchronous final state where rotational angular velocity is
related to, but not exactly, the average orbital angular velocity
\citep{Hut81},

\beq{eq:nsync}
n\sub{sync}\equiv\frac{\Omega\sub{sync}}{\omega}=
\frac{1+\frac{15}{2}e^{2}+\frac{45}{8}e^{4}+\frac{5}{16}e^{6}}
     {(1-e^{2})^{\frac{3}{2}}(1+3e^{2}+\frac{3}{8}e^{4})}
\end{equation}

For binary eccentricities $e$ in the range $0-0.5$,
$1<n\sub{sync}<2.8$, implying that provided enough time, stellar
components of a binary will be locked in a rotational state where
their periods are close to that of the binary period. This result will
not depend on the mass loss history of each star.  Once locked, the
magnetic activity of the stars becomes frozen at a given value, which
in some cases could mimic the behavior of a star that is much younger
or older \citep{Mason13}.

The second exogenic effect removing AM from the convective layer is
mass loss via magnetized SWs.  Not only does the mass leaving the
surface of the star carry AM, but in magnetically active stars the
flow of charged particles is also coupled with the corotating magnetic
field, up to few to several tens of stellar radii.  This leads to a
loss of AM that is several orders of magnitude larger than that in
stars without a convective envelope \citep{Schatzman62}.

Although a successful, comprehensive, and self-consistent model of
magnetized SW AM loss has not yet been developed, a semi-empirical
formula, developed by \citet{Kawaler88} and modified by
\citet{Chaboyer95}, succeeds at producing results consistent with
observations for a large range of stellar masses and ages.
Accordingly, the torque exerted by the magnetized SW scales with
angular velocity $\Omega$, stellar mass $M_\star$, and radius
$R_\star$, following

\beq{eq:TauML}
\Tau_{\rm SW}=K_{\rm C}\,\Omega\;{\rm min}(\Omega,\Omega_{\rm sat})^2\,
\left(\frac{R_\star}{R_\odot}\right)^{1/2}\,
\left(\frac{M_\star}{M_\odot}\right)^{-1/2}
\eeq

$K_C$ and $\Omega_{\rm sat}$ are two free parameters that can be
adjusted to reproduce the scale of observed rotational rates and their
distribution, respectively.  In order to reduce the number of free
parameters in our comprehensive model, $\Omega_{\rm sat}$ is
calculated following the scaling relationship

\beq{eq:OmegaSat}
\Omega_{\rm sat}=\Omega_\odot \left(\frac{\tau_{{\rm
    conv},\odot}}{\tau_{\rm
    conv}}\right)
\eeq

\noindent where $\tau_{\rm conv}$ is the convective overturn time (see
Section \ref{subsubsec:XUV}).  This scaling relationship has proven to
be useful in reproducing the distribution of rotation periods for
low-mass field stars \citep{Irwin11}.

More recently, success in modeling the rotational rate distribution of
stars in young clusters was achieved \citep{Matt12,Gallet13} using a
semianalytical fit of MHD simulations to the original model of
\citet{Kawaler88} along with a recent model for stellar mass loss
\citep{Cranmer11}. However, its application is still restricted to
stars with mass nearly identical to the Sun.  Our interest here is in
stars ranging from 0.2 to 1.05 $\Msun$; see Table 1.  A comparison
between the results obtained with the semi-empirical formula used here
(Eq. \ref{eq:TauML}) and those obtained with the formula by
\citet{Matt12} is left for a future work, but it is not expected to
alter the major conclusions of the current work.

As the convective envelope loses AM, a difference in angular velocity
between the envelope and the radiative core is established.  A variety
of physical processes, such as magnetic fields, hydrodynamical
instabilities, and gravity waves, may redistribute AM in the stellar
interior, reducing this difference with time.  To model AM
redistribution we follow, as is customary, the prescription by
\citet{MacGregorBrenner91}, by estimating the mutual torque due to the
difference in rotation rates as given by

\beq{eq:TauDJ}
\left|\Tau_{\rm \Delta J}\right|=\frac{\Delta J}{\tau_{\rm ce}}
\eeq
 
The AM exchanged is $\Delta J=(I_{\rm conv} J_{\rm core} - I_{\rm
  core} J_{\rm conv})/I_{\rm tot}$, and $\tau_{\rm ce}$ is a free
parameter called the core-envelope coupling timescale.

In summary, the AM of the convective envelope and the radiative core
\hl{of each star in the binary} evolve following the equations

\begin{eqnarray}
\frac{dJ_{\rm conv}}{dt} & = & \Tau_{\rm tid} + \Tau_{\rm SW} + \Tau_{\Delta J} \\
\frac{dJ_{\rm rad}}{dt} & = & - \Tau_{\Delta J}.
\end{eqnarray}

The moment of inertia of both the convective envelope and the
radiative core evolves in time and affects the rotational history
\hl{of the whole star}.  PMS evolution is especially important \hl{in
  this respect}.  During PMS, stars contract the most. A pragmatic
approach to account for this early contraction is to integrate the
angular velocity instead of the AM using

\beq{eq:OmegaEOM}
\frac{d\Omega_i}{dt}=\frac{1}{I_i}\frac{dJ_i}{dt}-\Omega_i\frac{dI_i}{dt}
\eeq

\noindent where the term $-\Omega_i(dI_i/dt)$ accounts for the
contraction (expansion) of the respective layer.  \hls{It is important
  to stress that although the larger changes in $I_i$ happen during
  the PMS, we are also integrating Eq. (\ref{eq:OmegaEOM}) during the
  main sequence.}

From a purely dynamical point of view this contraction term is
associated with two discrepancies.  The rapid contraction of the star
during the first stages of stellar evolution is compounded with the
gain of specific AM by accreting gas. Angular acceleration of the star
may continue even to near breakup velocities ($P_{\rm rot}\sim 0.1$
days).  Observations, however, show that the rotation rates of PMS
stars are orders of magnitude slower (with periods of $2-10$ days).
Although still debated, the solution to this discrepancy may rest on
the assumption that stars transfer this AM excess to their accretion
disk.  This disk locking persists for a timescale $\tau_{\rm disk}\sim
2-5$ Myr, ending with disk dissipation.

The second discrepancy arises when considering the value and sign of
the contraction term for the radiative core.  In stellar evolution
simulations, the core suddenly forms at around $10-100$ Myr and grows
to its final size over a time an order of magnitude smaller.  As a
result, the moment of inertia of the core increases rapidly,
potentially producing a sudden rotational breaking of this layer.
Even after considering the exchange of AM during the growth of the
core, as is done in \citet{Allain98}, this sudden braking still
occurs.  Although unobservable, a sudden deceleration of the radiative
core would produce a signature in the evolution of the envelope and
would create a discrepancy with observations.

To be consistent with previous results, we assume that the contraction
term for the core in Eq. \ref{eq:OmegaEOM} is the same as that of the
convective envelope.  An investigation of how this apparent
deceleration in the radiative core seems to be prevented is a critical
matter in need of further research.

\hl{For each system studied here, numerical integrations of
  Eqs. (\ref{eq:OmegaEOM}) were performed for both stars in the binary
  including their mutual tidal interaction.  For that purpose we take
  as inputs the values of key stellar structure properties as a
  function of time (e.g. apsidal motion constants, moments of inertia,
  and convective zone thickness), from stellar evolutionary models
  \citep{Bressan12,Bressan13} as well as from interior structure
  models (\href{}{Baraffe, 2014, personal communication})}.

%%%%%%%%%%%%%%%%%%%%%%%%%%%%%%%%%%%%%%%%%%%%%%%%%%%%%%%%%%%%%%%%%%%%%%%%%%%%%%%%
\subsection{Magnetic Activity}
\label{subsec:MagneticActivity}
%%%%%%%%%%%%%%%%%%%%%%%%%%%%%%%%%%%%%%%%%%%%%%%%%%%%%%%%%%%%%%%%%%%%%%%%%%%%%%%%

The ultimate purpose of calculating instantaneous stellar rotational
periods is to calculate the magnetic activity as a function of time.
Magnetic phenomena in the atmospheres of cool stars are driven by the
action of a dynamo in their convective zones that in turn is
responsible for producing SWs and high-energy radiation (XUV
emission).

Progress in understanding and characterizing these processes in the
Sun, allowed S. Cranmer and S. Saar to develop a self-consistent model
of SW acceleration applicable to other cool stars \citep{Cranmer11}.
Their model accurately predicts stellar activity observables such as
the average photospheric magnetic field strength and mass loss rates
as a function of several basic stellar properties such as mass,
radius, luminosity, and rotational period.

Their publicly available routine {\tt BOREAS} is applied here in order
to calculate the instantaneous mass loss rate of each star in the
binary system.  For details on the input physics, assumptions, and
free parameters used by the models of Cranmer and Saar, refer to
\citet{Cranmer11}.

\subsubsection{SW}
\label{subsec:StellarWind}

Armed with the mass loss rate $\dot M$, the properties of the SW are
calculated in a straightforward manner.  We assume that the wind
reaches its terminal velocity $v_{\rm SW}$ at a distance $r_t$ from
the stellar center, being equal to the local escape velocity:

\beq{eq:vsw}
v_{\rm SW}=\sqrt{\frac{2 G M_\star}{r_t}}
\eeq

\hl{A distance} $r_t=2.5 R_\star$ is assumed in order to reproduce the
current average solar wind condition as observed at Earth distance.

By mass conservation, the particle mean density as measured at a
distance $r$ from the star is given by

\beq{eq:nsw}
n_{\rm SW}=\frac{\dot M}{\mu m_p 4\pi r^2 v_{\rm SW}}
\eeq

\noindent where $\mu$ is the mean molecular weight of the wind
particles and $m_p$ is the proton mass.  For a fully magnetized SW
$\mu=0.6$ \citep{Cranmer08, Cranmer11}.

This stream of charged particles continuously impacts the atmosphere
of planets, at a rate given by

\beq{eq:FSW}
F_{\rm SW}=n_{\rm SW} v_{\rm SW}.
\eeq

At the distance of the HZ, the wind is supersonic and has a negligible
thermal pressure.  As a result, it exerts a dynamic pressure over
planetary magnetospheres given by

\beq{eq:FSW}
P_{\rm SW}=\mu m_p n_{\rm SW} (v_{\rm SW}^2+v_{\rm orb}^2)
\eeq

\noindent where $v\sub{orb}$ is the orbital velocity of the planet.

The Cranmer \& Saar model has been developed in order to describe
activity in single-stars.  The previous assumptions concerning how to
convert mass-loss rates into SW properties, may work well in
describing SWs for moderately spaced binaries, as well as isolated
stars, but the applicability of these models to binaries has not been
tested.  Additionally, other effects could arise, modifying the
underlying assumptions on which this model is built.  For example, a
star could illuminate the atmosphere of the companion, thereby
modifying the vertical atmospheric structure. Coronal X-ray heating
may play a critical role in some cases. The acceleration due to
gravity is also modified by the presence of a close companion.
Stellar wind shocks could also arise, modifying the simple model
described herein.  We assume that the acceleration mechanism working
in single stars provides a valid constraint for stars in the
moderately separated binaries under consideration.

At distances large enough from the binary or in very disparate
binaries ($q\ll 1$), it is reasonable to assume that the SW
flux and dynamic pressure are simply the sum of the quantities produced
by each star, i.e.

\beq{eq:FSW_Bin}
F_{\rm SW,bin}\approx F_{\rm SW,1}+F_{\rm SW,2}
\eeq

\beq{eq:PSW_Bin}
P_{\rm SW,bin}\approx P_{\rm SW,1}+P_{\rm SW,2}.
\eeq

For close orbiting planets, these simple assumptions are expected to
break down under realistic conditions in some binaries.  \hl{Recently,
  the conditions under which the interaction between the winds of
  binaries could strongly interact, rendering this approximation
  imprecise, have been studied using MHD simulations.  The results of
  these simulations suggest that even in an extreme case of solar-mass
  twins, the SW flow is only strongly modified at close circumbinary
  distances. Within the BHZ the simple approximation given in
  Eq. (\ref{eq:FSW_Bin}) is applicable for the purposes of studying
  the plasma environment of potentially habitable planets.}

\subsubsection{XUV Emission}
\label{subsubsec:XUV}

One of the most detrimental effects that planets face is early stellar
activity.  During the early history of planets, the high-energy
radiation flux far exceeds that which they receive later on.  In the
case of Earth for example, it has been estimated \citep{Guinan09} that
4.5 Gyr ago the X-rays flux was 10-100 times larger than present Earth
level (hereafter PEL) owing to solar coronal activity.

The effect that such flux has over time on planetary atmospheres
ranges from simply a modification of the upper atmospheric chemistry
\citep{Tian08, Segura10, Hu12, Hu13, Hu14} to rapid photoevaporation
via induced heating of the exosphere (\citealt{Lammer03};
\citealt[p. 399]{Lammer09d}; \citealt{Lammer10};
\citealt{Zendejas10}). Both phenomena could have very detrimental
effects on planetary habitability. Estimating the flux of high-energy
radiation is thus an essential element in the assessment of
habitability of any planetary environment.

The emission of X-Ray and EUV radiation (\hl{1-91.2 nm}) and their
dependence on stellar age and rotational period have been extensively
studied in the case of single-stars.  A strong anticorrelation between
age and XUV luminosity has been observed \citep{Ribas05, Penz08a,
  Penz08}.

Binary stars follow a different set of rules. First, rotation and age
are not necessarily correlated, \hl{especially} when the binary period
is short (of the order of days) or if the binary eccentricity is
larger than about 0.1, owing to tidal torques.  Second, and as
explained before, strong magnetic interaction between the stellar
components in the first tens of Myr could drive rotational evolution
and hence activity toward values that are uncommon in single-stars.

\hl{In spite of the most common recipes used to estimate XUV levels as
  a function of stellar age (see, e.g., \citealt{Zuluaga12}), which
  will not be applicable in the case of binaries for the reasons given
  above, we use here an approach consistent with the methods used to
  calculate mass loss rates, i.e. using the activity proxies provided
  by the models of \citet{Cranmer11}.}

Using a catalog of $\sim$1600 stars whose rotation and X-ray
brightness have been measured, \citet{Vardavas05} and more recently
\citet{Wright13} showed that X-ray luminosity is strongly correlated
with the {\it Rossby number} (the ratio of inertial to the Coriolis
force in the convection zone).  The Rossby number is calculated as

\beq{eq:Ro}
Ro=P_{\rm rot}/\tau_{\rm conv}
\eeq

Here $P_{\rm rot}$ is the rotational period \hl{(which determines
  Coriolis forces)} and $\tau_{\rm conv}$ (the convective overturn
time) is the typical timescale for convective motions (inertial
forces).

\citet{Wright13} found the following empirical relationship:

\beq{eq:LX}
R_{\rm X}=\frac{L_{\rm X}}{L_{\rm bol}}=R_{\rm X,sat}\times\left\{
\begin{array}{ll}
1.0 & \rm{if}\;{\rm Ro}<{\rm Ro}_{\rm sat} \\
({\rm Ro}/{\rm Ro}_{\rm sat})^{-2.7}  & \rm{otherwise}
\end{array}
\right.
\eeq

Here ${\rm Ro}_{\rm sat}$ and $R_{\rm X,sat}$ are fitting constants
quantifying the Rossby number at which the emission is saturated and
the constant ratio of X-ray to bolometric luminosity observed at
saturation levels.  Using their catalog, \citet{Wright13} found that
(${\rm Ro}_{\rm sat}$,$R_{\rm X,sat}$) are between a high-level range
of (0.17,0.001075) and a low-level range of (0.09, 0.00051).

It is interesting to notice that at a given value of ${\rm Ro}$ the
X-ray luminosity could be predicted \hl{using the empirical law in
  Eq. (\ref{eq:LX})} with an uncertainty of a factor of 2 even among a
large range of ages, stellar masses, and rotational periods.

In order to incorporate this result into our model, we need to predict
the Rossby number of a star at a given stage of its rotational
evolution.  For that purpose, the routines of \citet{Cranmer11}
that provide the convective overturn times as a function of stellar
effective temperature are used.

%%%%%%%%%%%%%%%%%%%%%%%%%%%%%%%%%%%%%%%%%%%%%%%%%%%%%%%%%%%%%%%%%%%%%%%%%%%%%%%%
\section{The Solar Reference Model}
\label{sec:SolarSystem}
%%%%%%%%%%%%%%%%%%%%%%%%%%%%%%%%%%%%%%%%%%%%%%%%%%%%%%%%%%%%%%%%%%%%%%%%%%%%%%%%

\begin{figure}
\begin{center}
\includegraphics[width=80mm,angle=0]{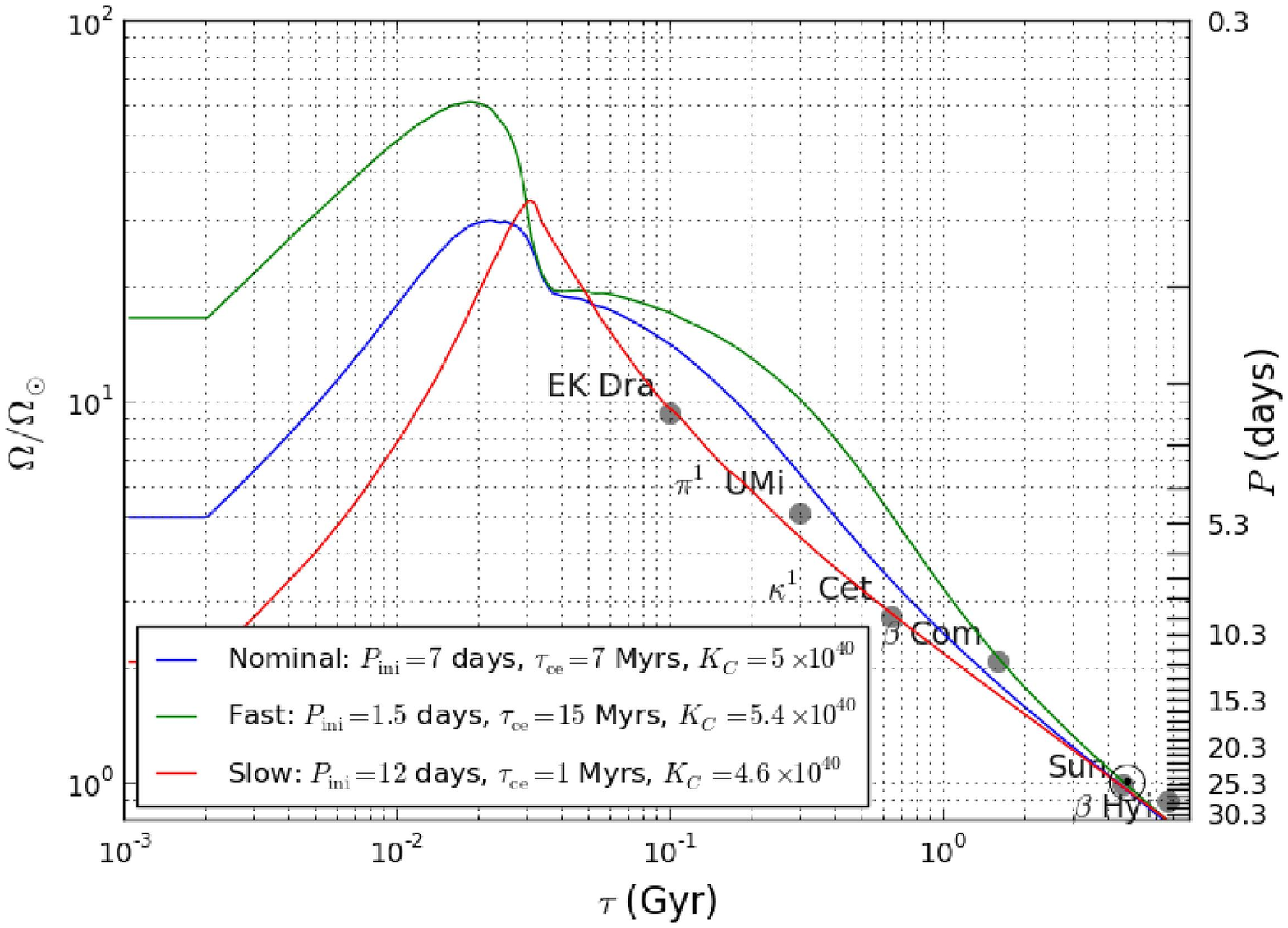}\\\vspace{0.3cm} %DONE
\includegraphics[width=75mm,angle=0]{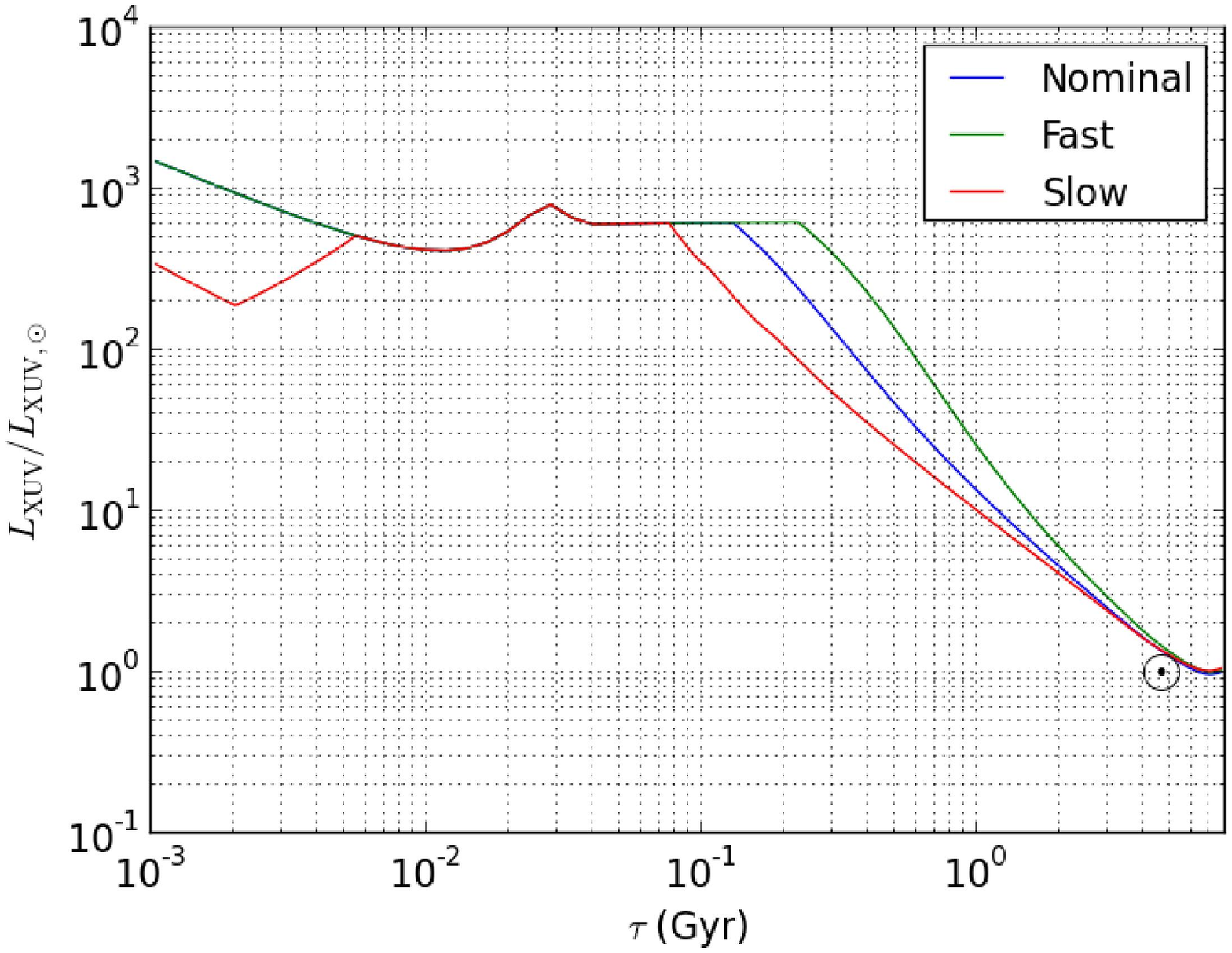}\\\vspace{0.3cm} %DONE
\includegraphics[width=75mm,angle=0]{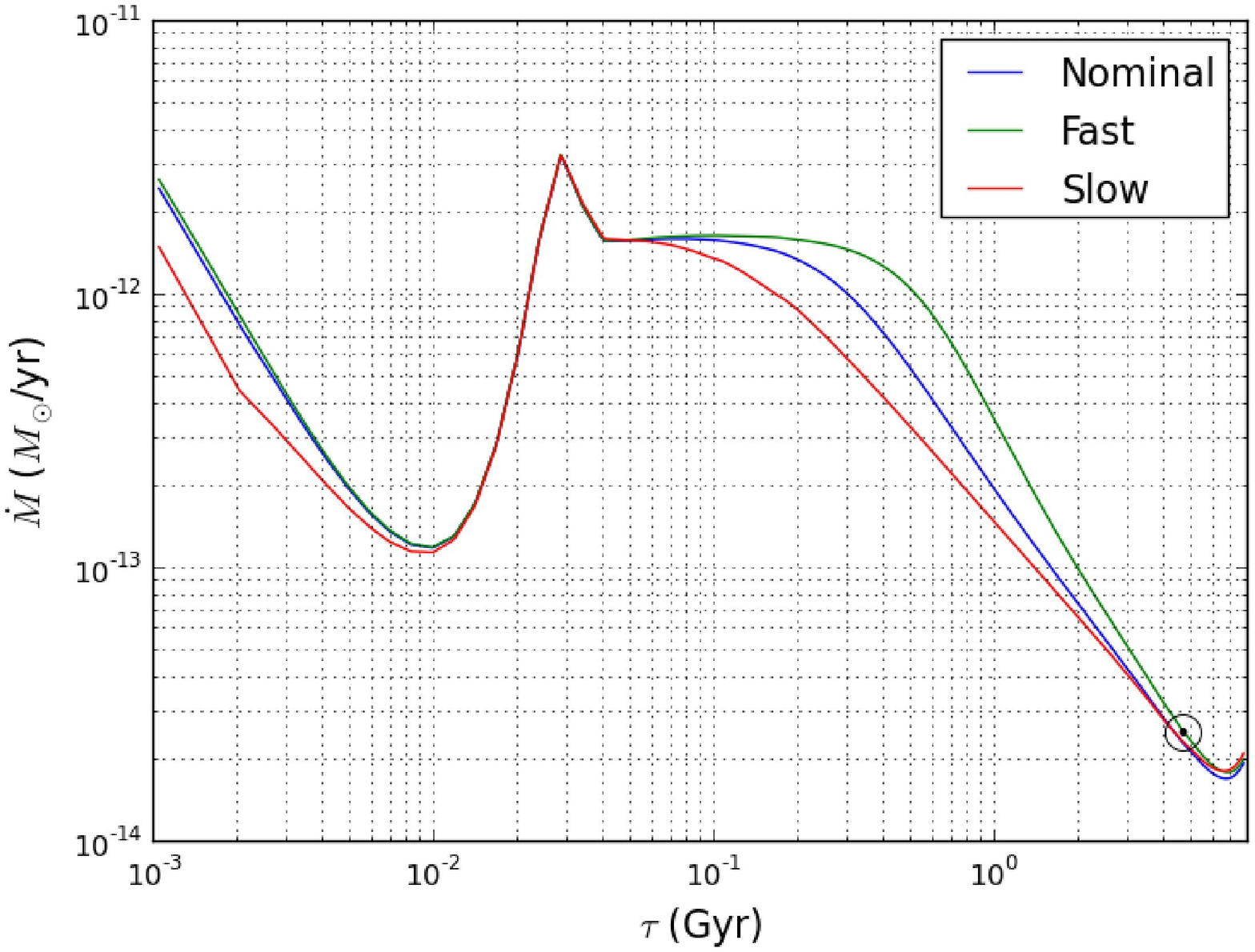} %DONE
\caption{Rotation and activity evolution of a solar-mass star in three
  different scenarios\hl{: a slow, nominal, and fast initially
    rotating Sun (see text)}.  Upper panel: rotational velocity as a
  function of time.  The measured rotations of five solar-like stars
  (masses between 1.01 and 1.10) {\bf and the Sun itself are} included
  for reference (\hl{the periods of rotation of the reference stars
    are taken from} \citealt{Ribas05}).  Middle panel: XUV luminosity
  as a function of time in terms of the present-day solar X-ray
  luminosity.  Lower panel: evolution of mass-loss
  rate.\label{fig:SolarReference}}
\end{center}
\end{figure}

In order to calibrate binary star models, we have calculated the
evolution of rotation and activity for an isolated solar-mass star.
The results are presented in Figure \ref{fig:SolarReference}.  We call
this the {\it Solar Reference Model} \hl{(SRM)}.

The free parameters of the model that were described in previous
sections have been chosen to reproduce the observed properties of the
Sun, namely, the period of rotation, mass loss rate, and XUV
luminosity at present times.

%FFFFFFFFFFFFFFFFFFFFFFFFFFFFFFFFFFFFFFFFFFFFFFFFFFFFFFFFFFFFFFFFFFFFFFFFFFFFFFFF
%FIGURE 3
%FFFFFFFFFFFFFFFFFFFFFFFFFFFFFFFFFFFFFFFFFFFFFFFFFFFFFFFFFFFFFFFFFFFFFFFFFFFFFFFF
%FFFFFFFFFFFFFFFFFFFFFFFFFFFFFFFFFFFFFFFFFFFFFFFFFFFFFFFFFFFFFFFFFFFFFFFFFFFFFFFF

Since activity and rotation of the Sun during the early phases of
solar system evolution are poorly constrained, three scenarios are
considered: (1) a nominal scenario (initial rotation period $P_{\rm
  rot,ini}=7$ days, $\tau_{\rm ce}=7$ Myr), (2) an initially
fast-rotating Sun ($P_{\rm rot,ini}=1.5$ days, $\tau_{\rm ce}=15$ Myr)
and (3) an initially slowly rotating sun ($P_{\rm rot,ini}=12$ days,
$\tau_{\rm ce}=1$ Myr).  In each case we use a disk locking time
$\tau_{\rm disk}=2$ Myr, in agreement with the estimations of the
solar nebula dissipation time \citep{Krot05}.  For each scenario the
value of the braking constant $K_C$ has been set so as to reproduce
the properties of the Sun at $\tau=4.56$ Gyrs (see legend in upper
panel of Figure \ref{fig:SolarReference}).

Using rotation periods, the mass-loss rate, calculated from the {\tt
  BOREAS} routine, as a function of time for each scenario is shown in
the lower panel of Figure \ref{fig:SolarReference}.  The XUV
luminosity has been calculated in each case using Eq. \ref{eq:LX} and
by assuming a critical Rossby number ${\rm Ro}_{\rm sat}=0.16$ and a
saturation X-ray luminosity of $R_{\rm X,sat}=7.4\times10^{-4}$.
These values are chosen in order to reproduce the observed X-ray
luminosity of the Sun.

With the calculated XUV luminosities and mass loss rates in hand, the
SW and XUV fluxes measured at the distance of Venus, Earth, and Mars
are determined.  From these fluxes, we calculate the integrated effect
on a planetary atmosphere:

\beq{eq:IntegratedFluxes}
\Phi_{\rm XUV,SW}(\tau;\tau_{\rm ini})=\int_{\tau_{\rm ini}}^{\tau}
F_{\rm XUV,SW}(t) dt
\eeq

%FFFFFFFFFFFFFFFFFFFFFFFFFFFFFFFFFFFFFFFFFFFFFFFFFFFFFFFFFFFFFFFFFFFFFFFFFFFFFFFF
%FIGURE 4
%FFFFFFFFFFFFFFFFFFFFFFFFFFFFFFFFFFFFFFFFFFFFFFFFFFFFFFFFFFFFFFFFFFFFFFFFFFFFFFFF
\begin{figure}
\begin{center}
\includegraphics[width=82mm,angle=0]{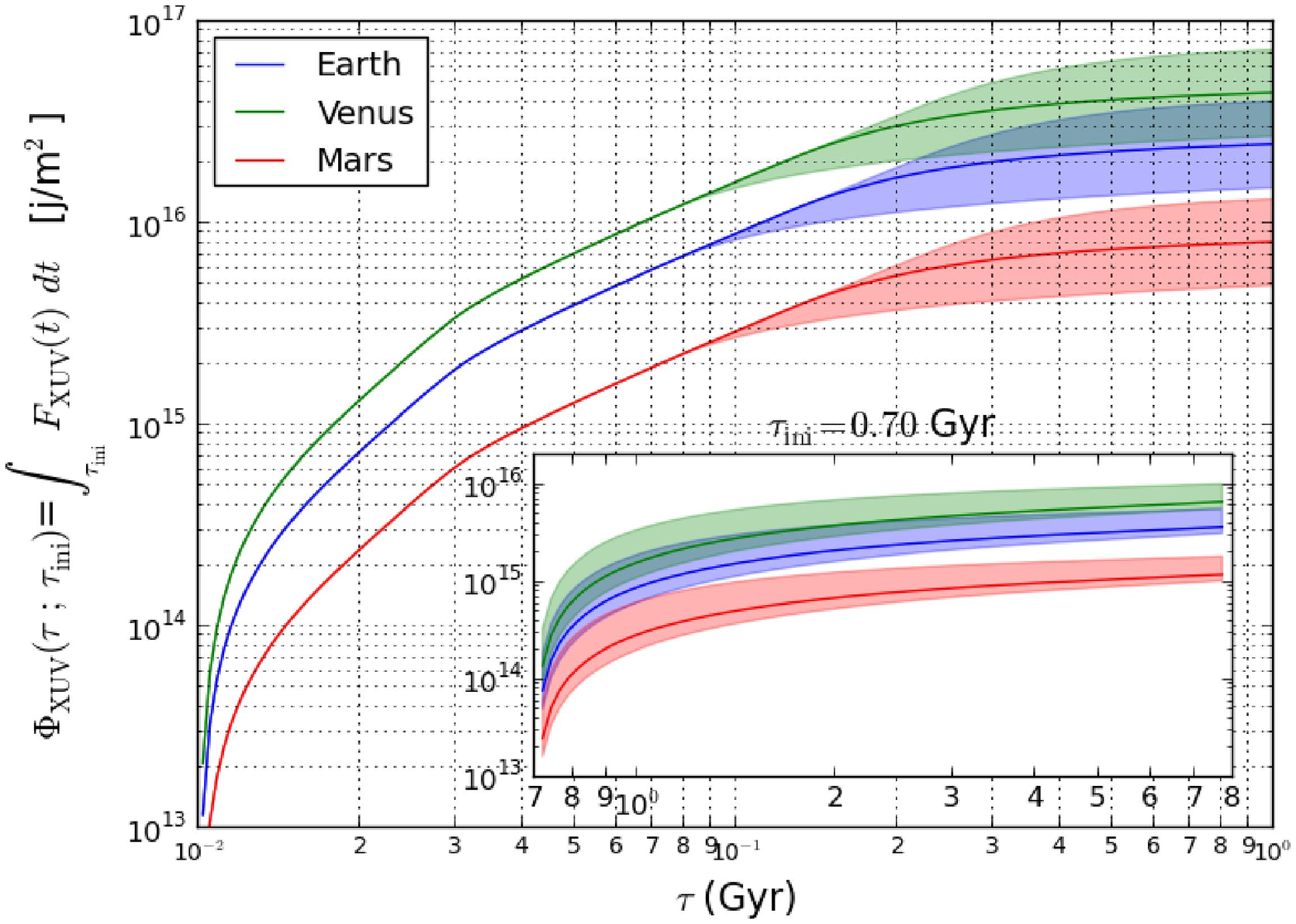}\\\vspace{0.5cm}%DONE
\includegraphics[width=82mm,angle=0]{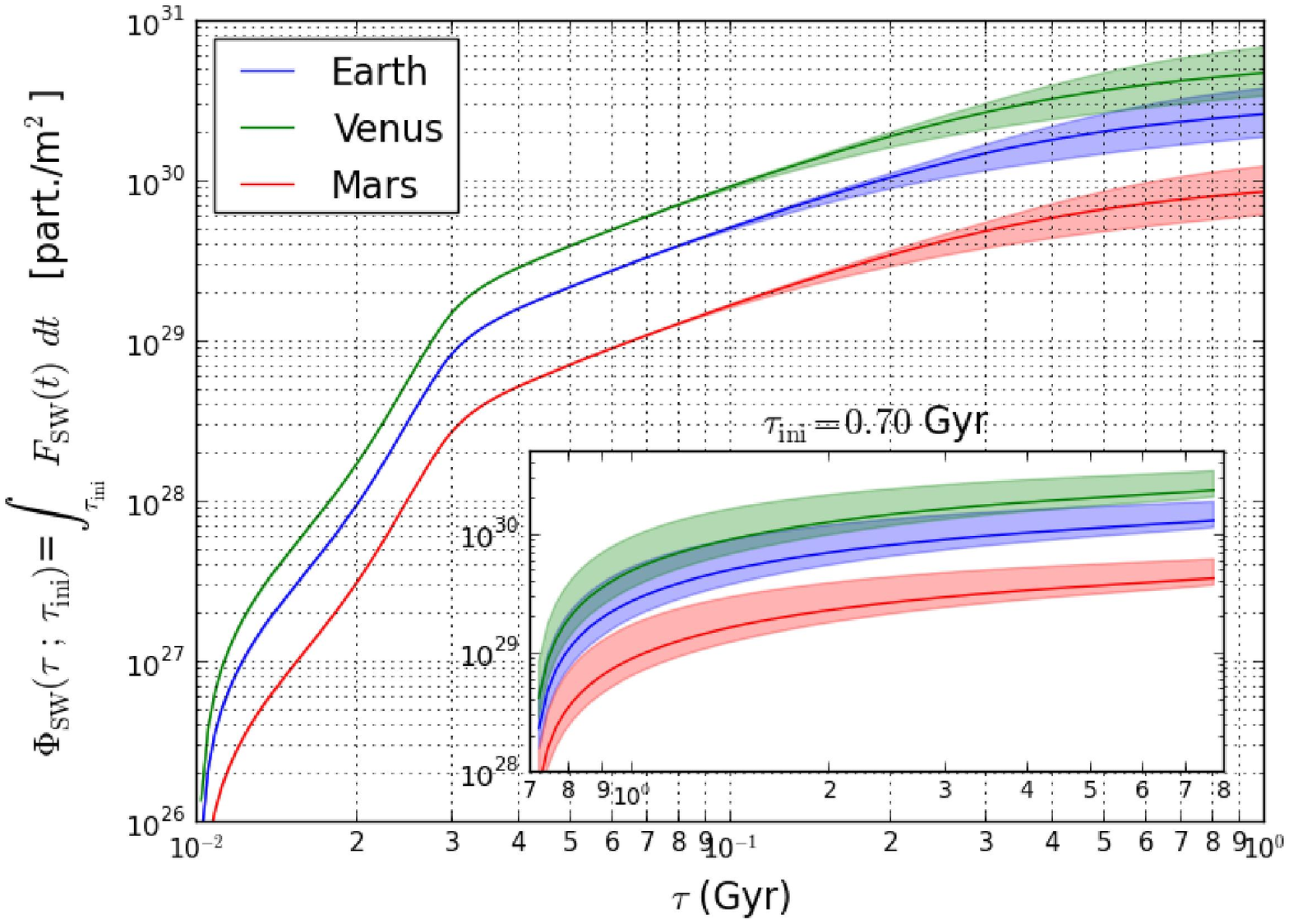}%DONE
\caption{Integrated XUV (upper panel) and SW (lower panel)
  fluxes as measured at Venus, Earth and Mars orbits. Integration is
  performed starting at 10 Myrs for the primary atmospheres and 700
  Myrs (inset panels) for secondary
  atmospheres.\label{fig:SolarReferenceFluxes}}
\end{center}
\end{figure}
%FFFFFFFFFFFFFFFFFFFFFFFFFFFFFFFFFFFFFFFFFFFFFFFFFFFFFFFFFFFFFFFFFFFFFFFFFFFFFFFF

Integrated values of XUV and SW fluxes are shown in Figure
\ref{fig:SolarReferenceFluxes}.  Integrated fluxes rapidly reach a
maximum around $200-400$ Myrs in the case of SW and $\sim$1 Gyr in the
case of XUV radiation.  At about those times, \hls{any hypothetical}
primitive planetary atmosphere \hls{(protoatmosphere)} has received
most of the incoming flux that it will ever receive.

If we assume that a \hls{``secondary''} atmosphere is degassed or left
over from the formation process \hl{(see, e.g., \citealt{Lammer13} and
  \citealt{Lammer13b} and references therein)}, we can also calculate
the cumulative effect that SW and XUV radiation has on these secondary
envelopes. In the inset panels of Figure
\ref{fig:SolarReferenceFluxes}, we plot the integrated flux starting
at 700 Myrs (around 3.8 Gyrs before present) and up to 8 Gyrs.  Again,
most of the integrated effects reach an asymptotic value after
$200-300$ Myrs (around 1 Gyr after formation).

Values of the integrated particle and photon fluxes can be converted
into total atmospheric mass loss from unmagnetized planets.  However,
converting one quantity into the other is nontrivial given the
complexity of the nonthermal and thermal mass loss processes involved.
For present purposes, first-order estimations can be obtained with the
phenomenological formalisms of \citet{Zendejas10} (for atmospheric
stripping) and \citet{SanzForcada11} (for XUV-driven erosion).

Using the asymptotic values of the integrated fluxes in Figure
\ref{fig:SolarReferenceFluxes}, a total solar-driven mass loss of the
order of $0.1-0.2$ bars for a \hls{hypothetical ``primary''
  atmosphere} of Mars is obtained.  Venus could have lost $1.5-3$ bars
from \hls{any protoatmosphere} as a result of the same mechanism.  In
both cases, \hls{and to be conservative}, we assumed that both planets
started with CO$_2$-rich atmospheres\footnote{\hls{The terms
    ``primary'' atmosphere, ``protoatmospheres,'' and ``secondary''
    atmospheres should not mislead the main aim of this work, which is
    to estimate the flux of particles and XUV radiation experiencied
    by planets in the early phases of stellar evolution.  For a
    thorough discussion regarding the mass, composition, and
    evolutionary phases of primitive atmospheres around Earth-like
    planets, please refer to \citet{Lammer13} and references
    therein.}}

For secondary atmospheres \hls{(i.e. any atmosphere remaining or
  degassed after several hundreds of Myr)}, we estimate that Mars and
Venus were subject to mass losses of the order of $0.06-0.1$ bars and
$0.8-1.5$ bars, respectively.  These numbers are in agreement with
previous estimations (see, e.g., \citealt{Kulikov06}).

In the case of Mars, which currently has a 6 mbar atmosphere, an
estimated mass loss of $\sim$100 mbar is \hls{fairly} compatible with
an early lost of ``habitable'' conditions \hls{(the presence of
  surface water)}.  Venus, having a \hls{present} $\sim$100 bar
atmosphere, seems to have barely lost 1\% of its mass via direct
stripping from solar wind.

If, on the other hand, we assume that terrestrial planets started with
massive hydrogen/helium envelopes (see \citealt{Lammer13} and
references therein), subject to photoevaporation from the absorption
of XUV radiation, we estimate (from the integrated XUV fluxes) that
the protoatmosphere of Venus could have lost 0.3-1\% of its total mass
in the first 300 Myrs.  This is of the same order as the mass a planet
with its size could have accumulated from the planetary nebula or from
other early outgassing processes \citep{ElkinsTantonSeager08}.

In summary, and recognizing that other complex endogenous and
exogenous factors are involved in the evolution of the atmosphere of
terrestrial planets and moons (vulcanism, degassing, impacts, etc.),
integrated SW and XUV fluxes calculated for the reference solar model
as well as for the {\it KBHZ sample}, \hls{are fairly compatible with
  the state of affairs in the solar system.  In consequence, our model
  can be used to} place first-order constraints on the critical level
of stellar aggression that planets \hls{around single and binary
  stars} could potentially withstand, \hls{before losing their water
  inventory and becoming uninhabitable}.

%%%%%%%%%%%%%%%%%%%%%%%%%%%%%%%%%%%%%%%%%%%%%%%%%%%%%%%%%%%%%%%%%%%%%%%%%%%%%%%%
\section{Rotational Evolution of {\it Kepler} Circumbinary Planets}
\label{sec:RotationalEvolutionKeplerStars}
%%%%%%%%%%%%%%%%%%%%%%%%%%%%%%%%%%%%%%%%%%%%%%%%%%%%%%%%%%%%%%%%%%%%%%%%%%%%%%%%

\begin{figure}
\begin{center}
\includegraphics[width=80mm,angle=0]{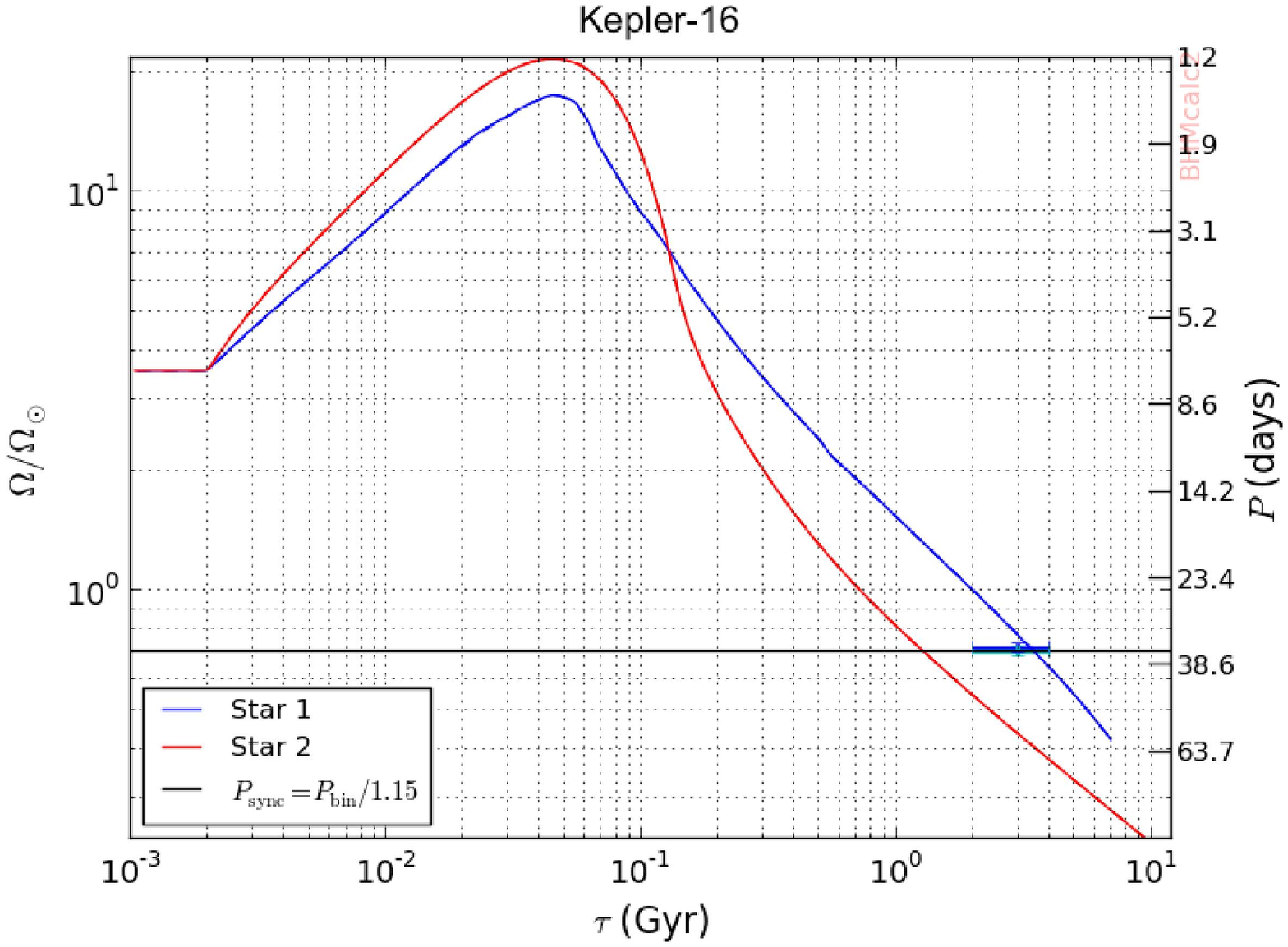}\\\vspace{0.25cm}
\includegraphics[width=80mm,angle=0]{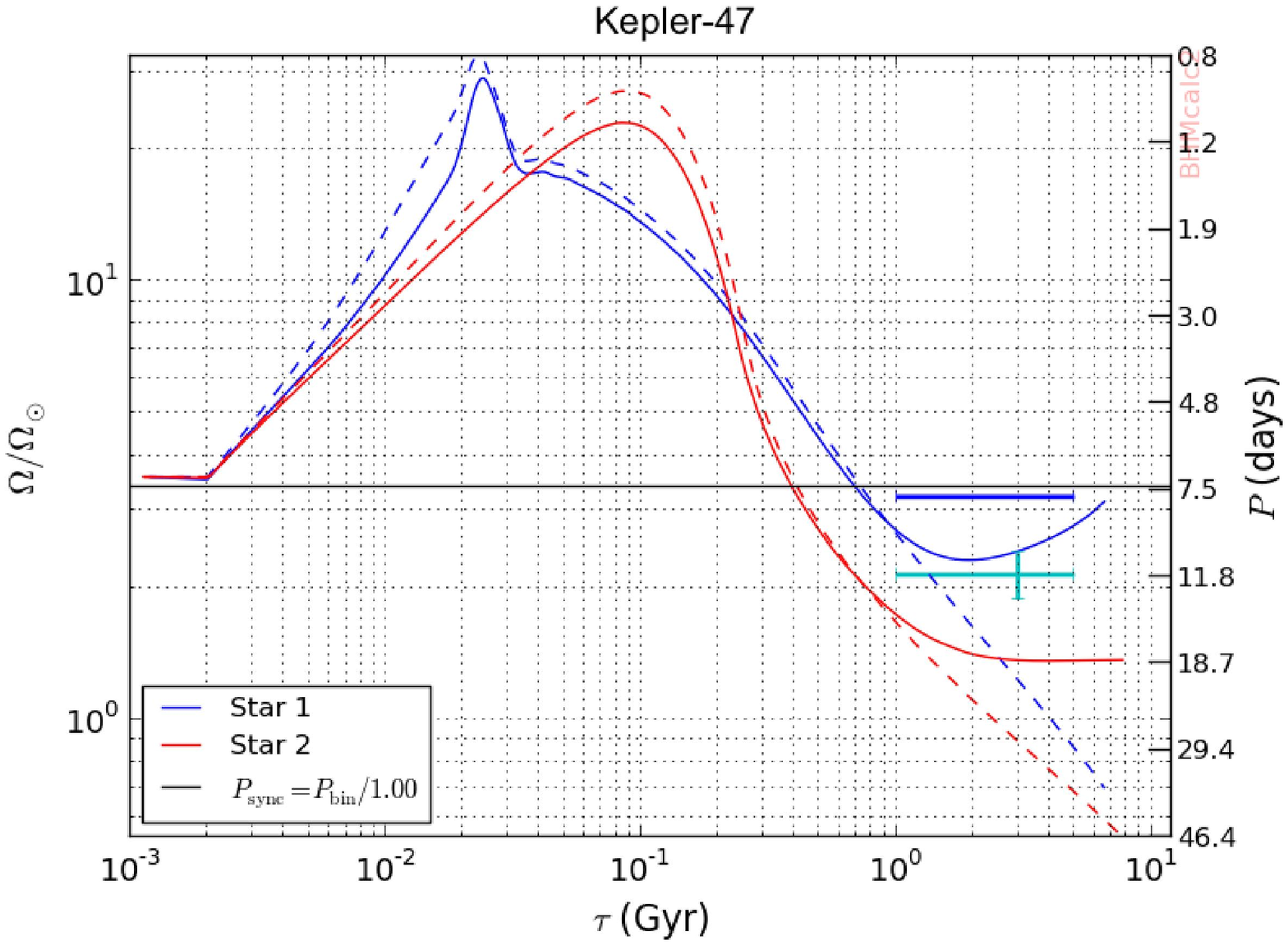}\\\vspace{0.25cm}
\includegraphics[width=80mm,angle=0]{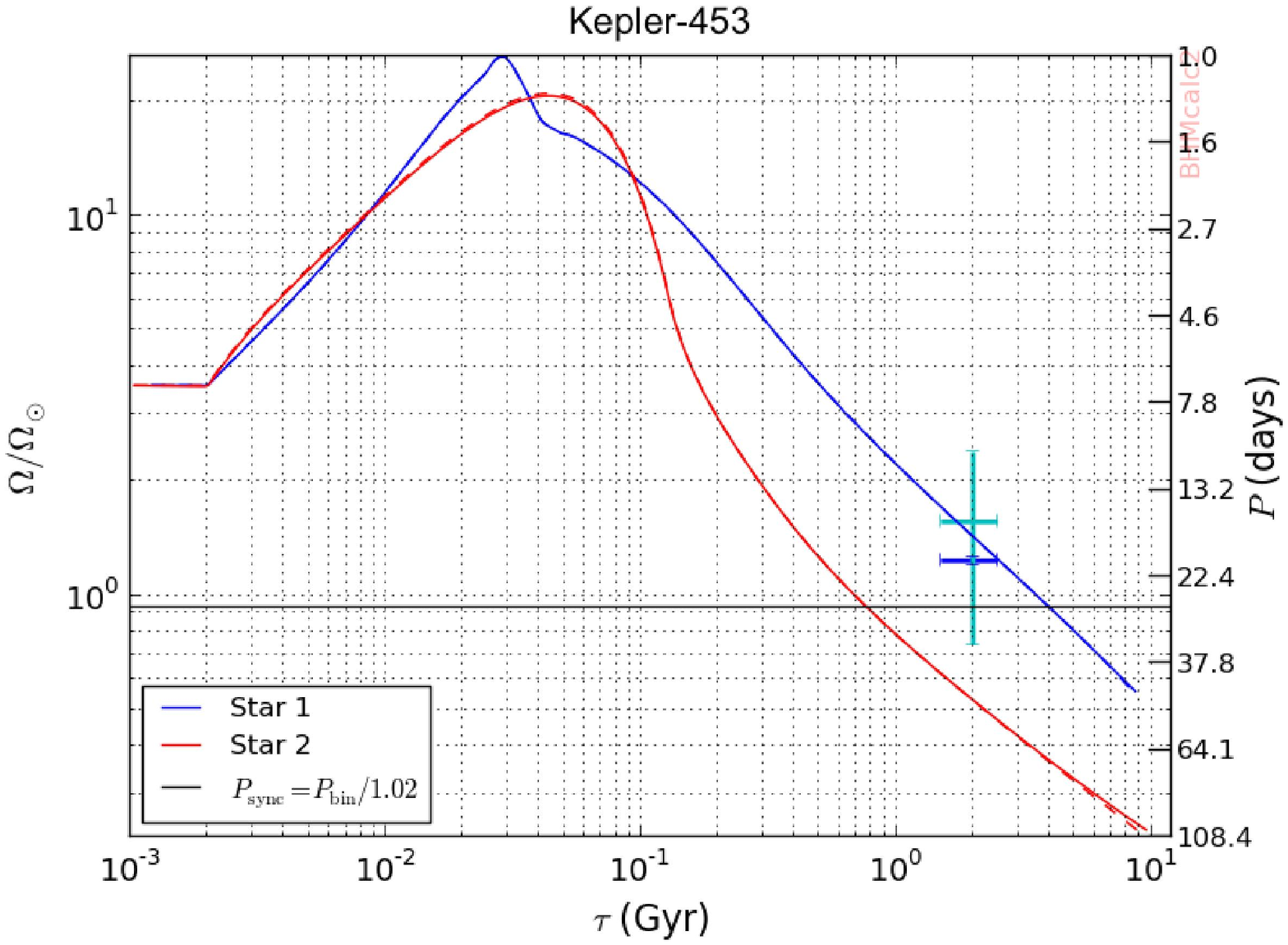}
\caption{Rotational evolution of each star in the {\it Kepler}
  binaries with an HZ planet, {\it KBHZ sample}.  \hl{Blue lines
    correspond to primaries and red to secondaries.  Solid lines show
    the evolution of rotation including the tidal interactions between
    the stars.  The dashed lines show the respective evolution if the
    stars were isolated. Independent age and rotation measurements
    (see Table 1) are shown with rotation from photometry (blue error
    bar) and spectroscopy (cyan error bar). The models developed here
    are in reasonable agreement with observed rotational periods (see
    text for discussion).}\label{fig:RotationEvolutionKBHZ}}
\end{center}
\end{figure}

The model developed and tested in previous sections is applied in order to
calculate the evolution of rotation and activity of the 
{\it KBHZ sample}.  Figure \ref{fig:RotationEvolutionKBHZ} shows rotation
rate as a function of time for each star in the three binary systems.

In each case, the equation of motion (Eq. \ref{eq:OmegaEOM}) including
the effect of tidal interaction is integrated.  \hl{For primary stars,
  we use the same parameters of the rotational evolution model
  ($\Omega\sub{sat}$, $K\sub{C}$, $P\sub{ini}$) as those of the solar
  reference nominal model.  Using the same parameters for the
  secondaries (all of them are low-mass stars $M<0.36 M_\odot$)
  produces anomalously low rotational rates at late times}.  In order
to be consistent with observed periods or rotation of field M dwarfs,
we use a much lower braking parameter, $K_C=10^{40}$ for companions.
\hl{With this adjustment the rotation periods} of Proxima Centauri and
Barnard's Star are both accurately predicted.

Although there are discrepancies between the rotation periods of the
primary star measured photometrically and spectroscopically (blue and
cyan error bars in Figure \ref{fig:RotationEvolutionKBHZ},
respectively), the model (blue line) predicts the observed values
reasonably well considering the large uncertainty box.

The observed period of Kepler-16A, $P_{\rm rot} = 35.1 (35.8)$ days,
is very close to that expected if the star is synchronized with the
binary orbit.  For the present eccentricity of the system the
pseudosynchronous period is $P_{\rm sync}=35.6$ days (see
Eq. \ref{eq:nsync}).  However, the tidal torque, even for the shortest
expected value of tidal dissipation, is too low to achieve
synchronization during the first couple of Gyrs \hl{(Nntice that the
  rotational evolution calculated without tidal interaction is shown,
  using dashed lines, in Figure \ref{fig:RotationEvolutionKBHZ}), for
  reference.}  Interestingly, the rotation evolution model also
predicts a rotation period very close to the pseudosynchronous period
for the estimated median age.  This may be considered a major
accomplishment for the proposed model.

The case of Kepler-47A is also very interesting. The period of
rotation measured photometrically, $P_{\rm rot,photo}=7.7$ days, is
close \hl{although} larger than the pseudo-syncronization period of
$P_{\rm sync}=7.4$ days.  Independently, a spectroscopically
determined velocity of rotation predicts a slightly larger rotation
period $P_{\rm rot,vel}=11.9$ days (provided that the inclination
angle is close to 90$^\circ$).  Stars in this system are close enough
to have their rotation periods substantially affected by tides.  Our
rotation evolution model predicts that within the estimated range of
ages for the system ($1-5$ Gyr), the rotation period will reach a
maximum of $\sim$10 days.  This value is in between the independently
measured periods. The primary star in the model is not synchronized as
early as \hl{first-order models suggest}.  \hl{Instead,} once all the
effects modifying the rotational rate during the PMS are taken into
account, tides become relevant only after several hundreds of Myr.

The case of \hls{Kepler-453} is more conventional, with tides having a
negligible effect on rotation during the relevant stellar evolution
stages.  The nominal model predicts a period of rotation slightly
larger than the photometrically measured one, but nearly identical to
the spectroscopically measured period. These three cases provide
evidence that rotational evolution is complex yet predictable by the
current model.

%FFFFFFFFFFFFFFFFFFFFFFFFFFFFFFFFFFFFFFFFFFFFFFFFFFFFFFFFFFFFFFFFFFFFFFFFFFFFFFFF
%FIGURE 5
%FFFFFFFFFFFFFFFFFFFFFFFFFFFFFFFFFFFFFFFFFFFFFFFFFFFFFFFFFFFFFFFFFFFFFFFFFFFFFFFF
%FFFFFFFFFFFFFFFFFFFFFFFFFFFFFFFFFFFFFFFFFFFFFFFFFFFFFFFFFFFFFFFFFFFFFFFFFFFFFFFF

%FFFFFFFFFFFFFFFFFFFFFFFFFFFFFFFFFFFFFFFFFFFFFFFFFFFFFFFFFFFFFFFFFFFFFFFFFFFFFFFF
%FIGURE 6
%FFFFFFFFFFFFFFFFFFFFFFFFFFFFFFFFFFFFFFFFFFFFFFFFFFFFFFFFFFFFFFFFFFFFFFFFFFFFFFFF
\begin{figure*}
\begin{center}
\includegraphics[width=75mm,angle=0]{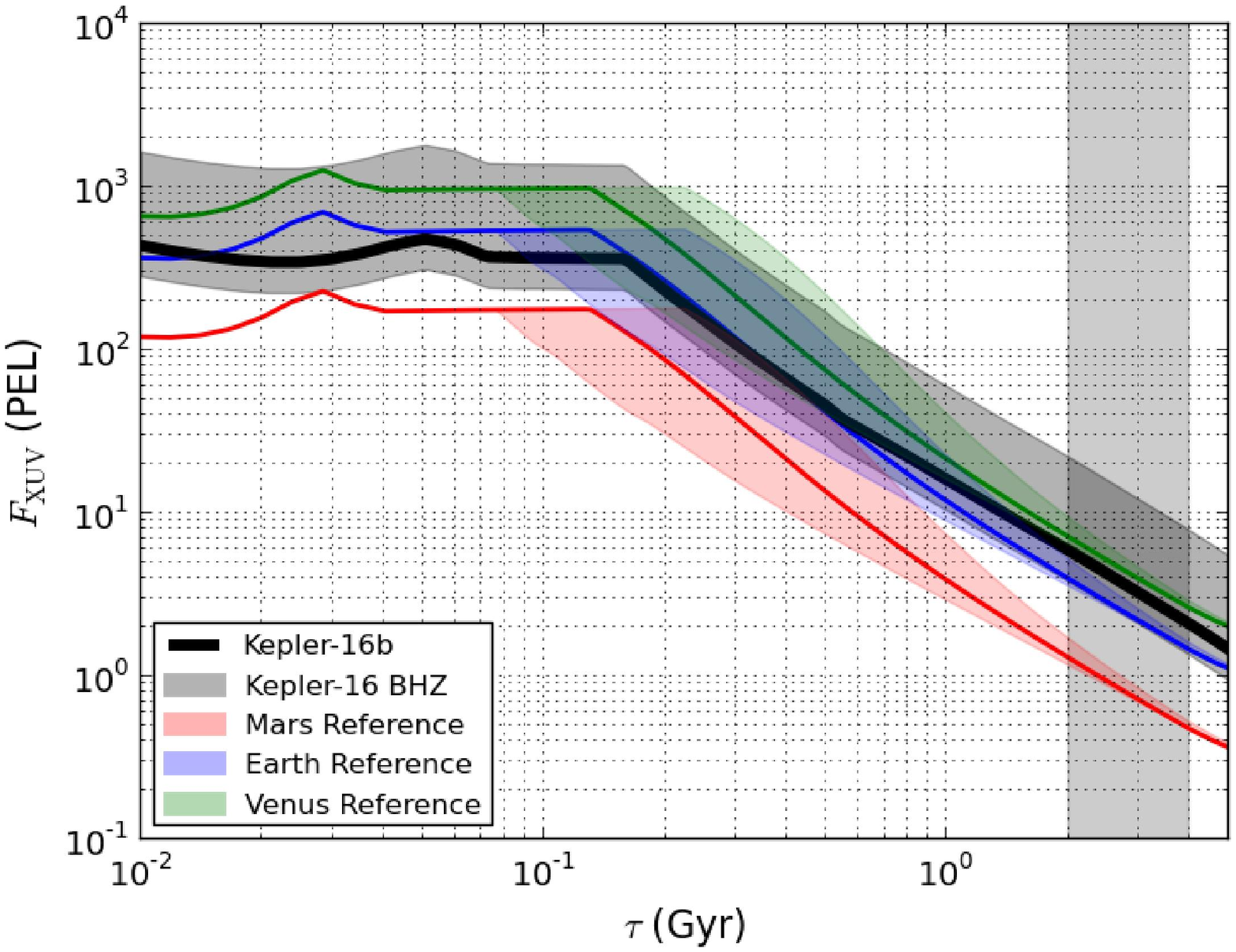}\hspace{0.5cm}%DONE
\includegraphics[width=75mm,angle=0]{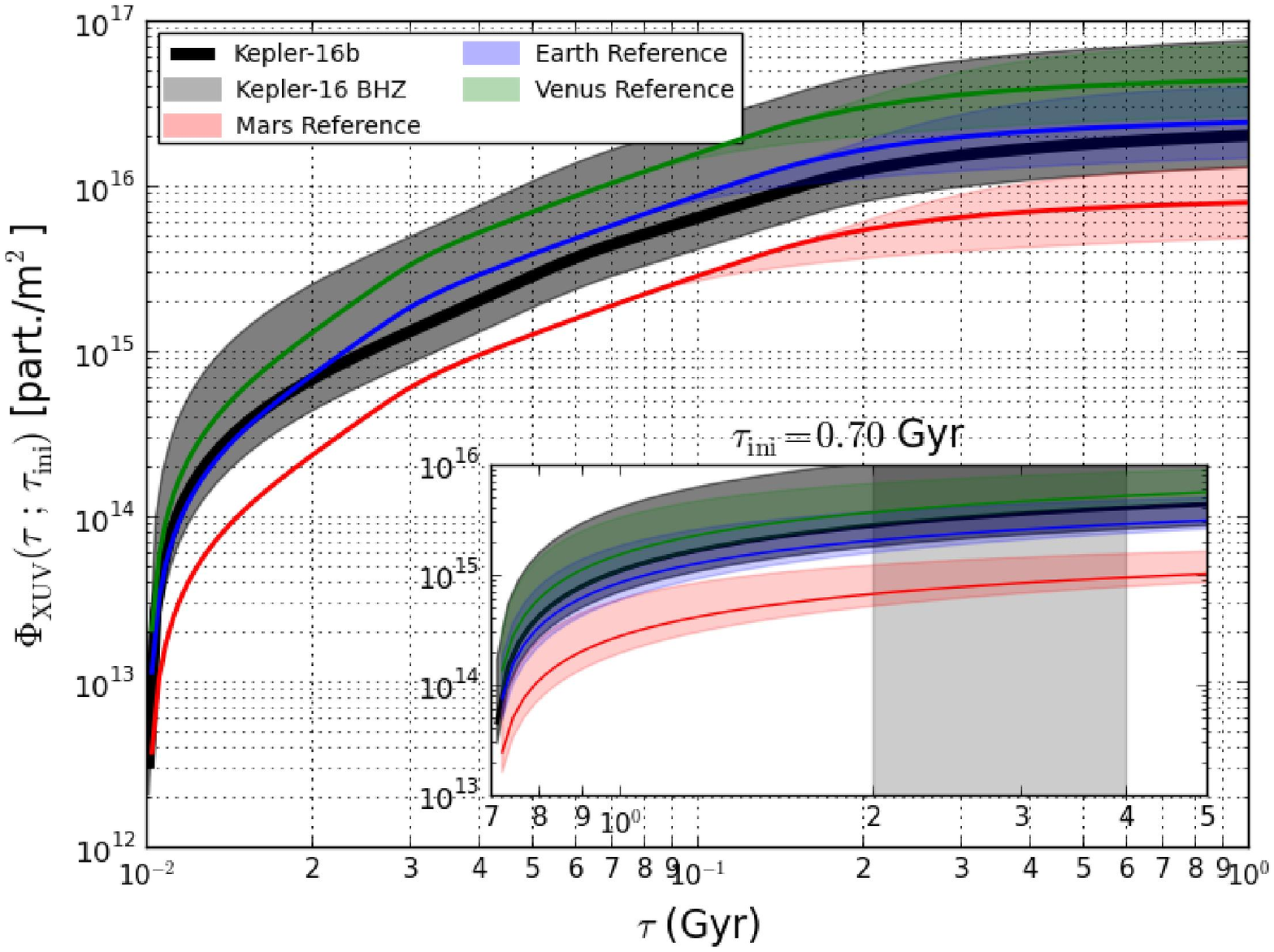}\\\vspace{0.2cm}%DONE
\includegraphics[width=75mm,angle=0]{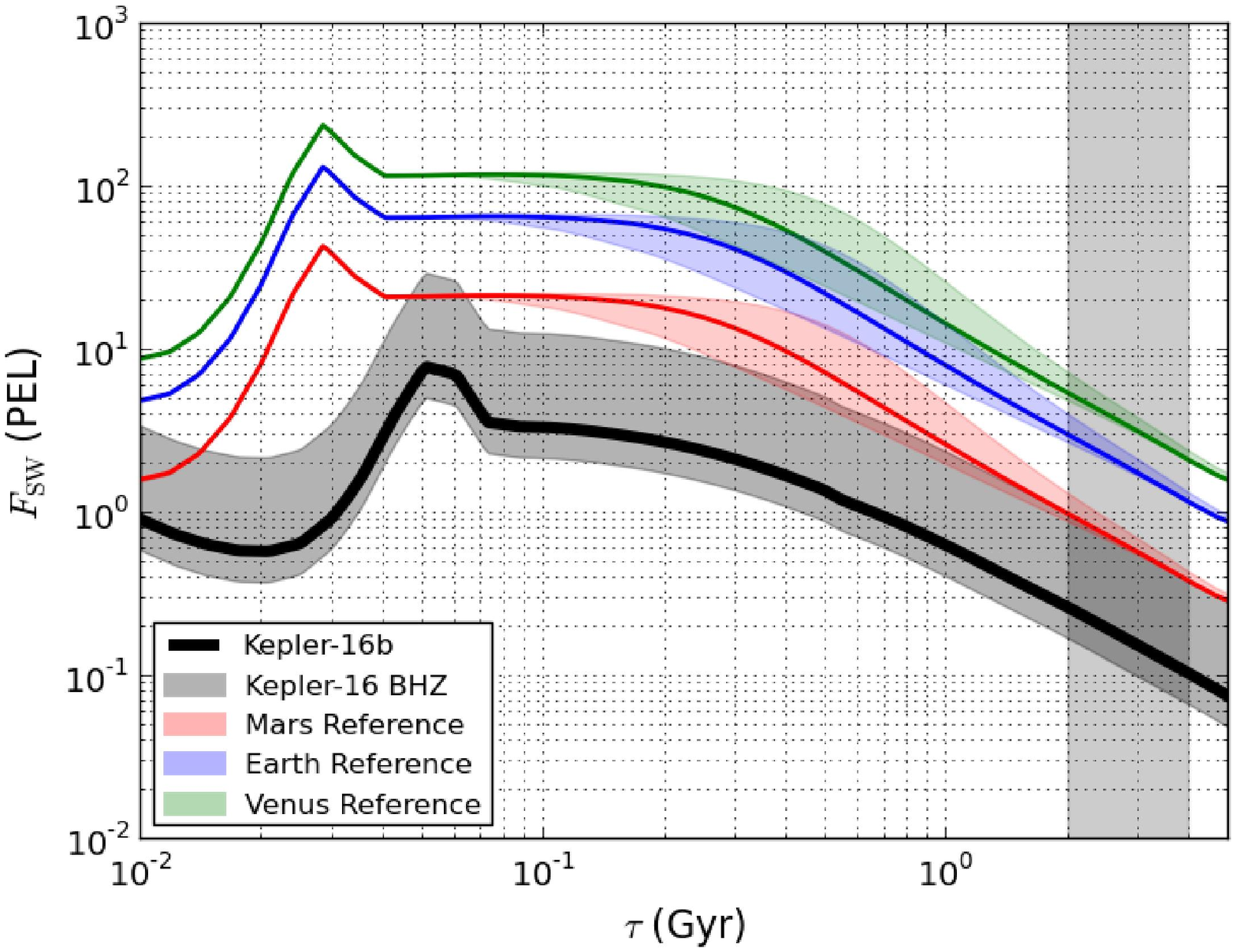}\hspace{0.5cm}%DONE
\includegraphics[width=75mm,angle=0]{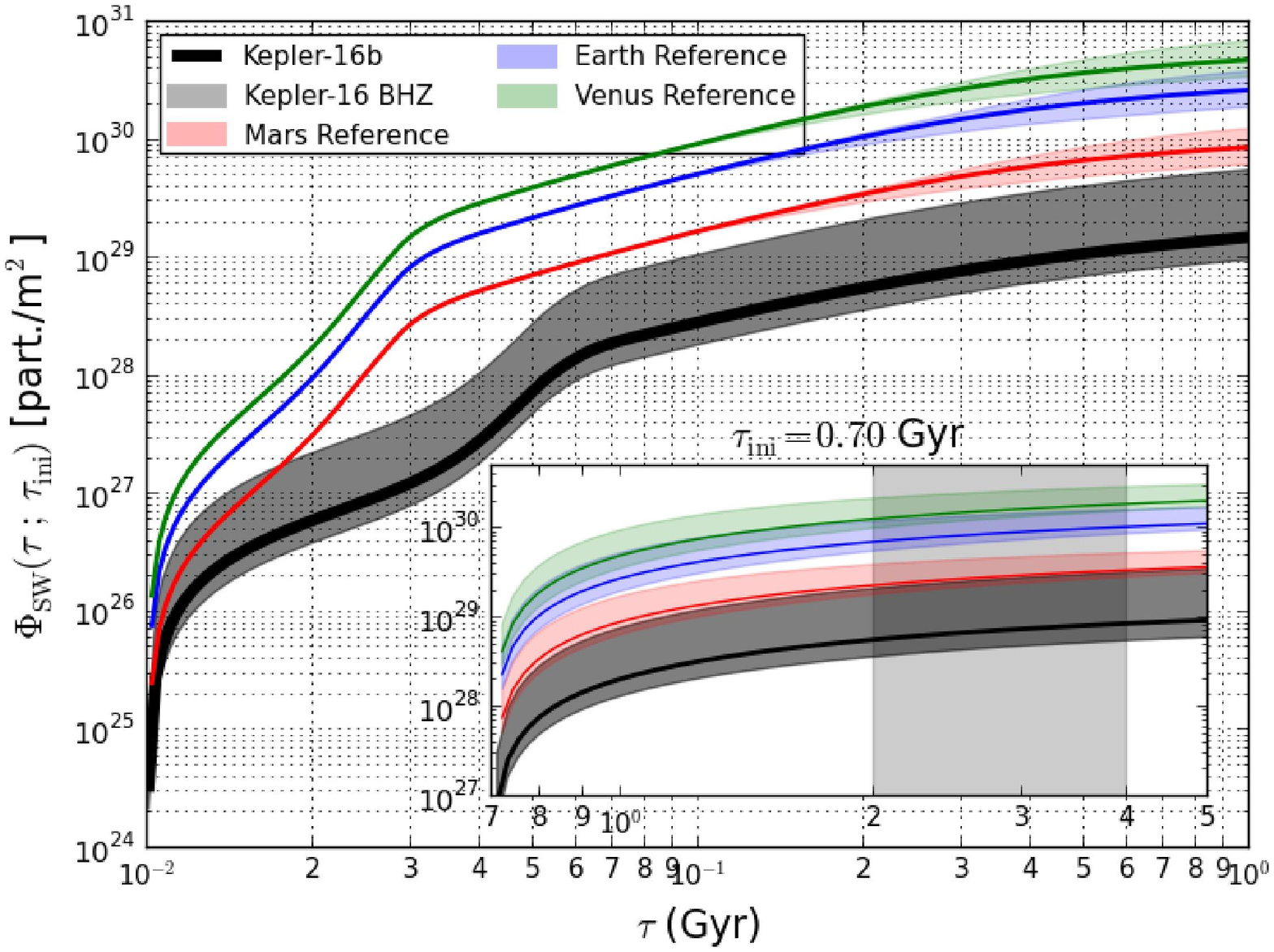}%DONE
\caption{SW and XUV flux evolution for
  Kepler-16.\label{fig:SWXUV-Kepler16} While harsh XUV conditions
  exist in the BHZ (top two panels), very favorable SW conditions are
  estimated (bottom two panels) to be better than in the solar
  system. A Mars-sized planet at the inner edge of the Kepler-16 BHZ
  resides in a plasma environment slightly better than Mars.}
\end{center}
\end{figure*}
%FFFFFFFFFFFFFFFFFFFFFFFFFFFFFFFFFFFFFFFFFFFFFFFFFFFFFFFFFFFFFFFFFFFFFFFFFFFFFFFF

%%%%%%%%%%%%%%%%%%%%%%%%%%%%%%%%%%%%%%%%%%%%%%%%%%%%%%%%%%%%%%%%%%%%%%%%%%%%%%%%
\section{Radiation and Plasma Environment of {\it Kepler} Circumbinary Planets}
\label{sec:RadiationPlasmaEnvironments}
%%%%%%%%%%%%%%%%%%%%%%%%%%%%%%%%%%%%%%%%%%%%%%%%%%%%%%%%%%%%%%%%%%%%%%%%%%%%%%%%

The focus of this paper is on the question of how the evolution of
rotation \hl{and activity} translates into a variable plasma and
radiation environment within the BHZ \hl{of three well-known
  circumbinary planets}.  Both effects impact the evolution of
potential planets and moons and their habitability.

Figures \ref{fig:SWXUV-Kepler16}-\ref{fig:SWXUV-Kepler47},
\ref{fig:SWXUV-KIC9632895} show the XUV radiation and SW fluxes as
calculated within the BHZ of the binaries (gray shaded region).
Instantaneous and integrated fluxes are shown.  In order to evaluate
the level of stellar aggression experienced around each {\it Kepler}
binary, we compare these results with those calculated for the {\it
  SRM} (solid thin lines).  \hl{We have compared fluxes around
  binaries with those for Venus, Earth, and Mars in the solar system.
  For each planet we calculate the range of flux experienced by those
  planets for two extreme solar activity evolution models (fast and
  slow initial rotation)}.

%%%%%%%%%%%%%%%%%%%%%%%%%%%%%%%%%%%%%%%%%%%%%%%%%%%%%%%%%%%%%%%%%%%%%%%%%%%%%%%%
\subsection{Kepler-16}
\label{subsec:Kepler16}
%%%%%%%%%%%%%%%%%%%%%%%%%%%%%%%%%%%%%%%%%%%%%%%%%%%%%%%%%%%%%%%%%%%%%%%%%%%%%%%%

\hl{The conditions around} Kepler-16 \hl{are shown in Figure
  \ref{fig:SWXUV-Kepler16}}. This binary system, which is composed of
K and M stars, has an XUV radiation environment harsher than that of
the solar system.  Most of the BHZ has been exposed to levels
comparable to, but mostly larger than, those of the solar system
planets.  \hl{Interestingly, the levels of XUV radiation received by
  Kepler-16b and any unobserved exomoon are almost identical to that
  received by the Earth in the solar system}.

On the other hand, particle fluxes in the BHZ of Kepler-16 are almost
one order of magnitude lower than that solar system HZ levels.  We
identify the source of this significant difference to be the wind
acceleration mechanism's dependence on stellar properties.

The mass loss rate depends on the amount of heat deposited in the
photosphere and chromosphere by the turbulent dissipation of
Alfv\'enic waves \citep{Cranmer11}.  This heat deposition depends,
among many other factors, on the third power of the wave amplitude
velocity that goes as $\rho^{-1/4}$, with $\rho$ the photospheric
density (see Eq. 14 in \citealt{Cranmer11}).  In smaller stars, with
larger photospheric densities, Alfv\'enic waves will propagate with
lower amplitudes, and as a result, less dissipated heat is available
to accelerate the wind.  In the case of Kepler-16, this, among other
less important factors, is responsible for a difference, from solar
values, of two orders of magnitude in the energy available for wind
acceleration.

The resulting effect is that, against all odds, the BHZ of Kepler-16
\hl{(and similar binaries)} exhibits enhanced conditions for habitable
low-mass planets \hl{or exomoons (if any)}.  We see in the \hl{bottom
  panels of } Figure \ref{fig:SWXUV-Kepler16} that even a Mars-sized
planet at the inner edge of the Kepler-16 BHZ will enjoy a plasma
environment slightly better than Mars.  \hl{Of course, orbital
  stability will limit, in this case, the very existence of such a
  planet, but it is still interesting to notice this effect}.
Moreover, if Kepler-16b or a similar planet has a Mars-sized exomoon,
then the SW appears to pose no serious threat to its atmosphere.

\hl{Is this effect a product of the fact that the planet is in a
  binary system?  No, actually.  Since the primary is just a bit less
  massive than solar and the companion has a very low-mass in a
  relatively wide orbit, stellar rotation periods have been barely
  affected by the presence of a companion.  Therefore, if we locate a
  planet in a scaled orbit around the primary, it will will enjoy
  similar advantageous conditions.}

%%%%%%%%%%%%%%%%%%%%%%%%%%%%%%%%%%%%%%%%%%%%%%%%%%%%%%%%%%%%%%%%%%%%%%%%%%%%%%%%
\subsection{Kepler-47}
\label{subsec:Kepler47}
%%%%%%%%%%%%%%%%%%%%%%%%%%%%%%%%%%%%%%%%%%%%%%%%%%%%%%%%%%%%%%%%%%%%%%%%%%%%%%%%

%FFFFFFFFFFFFFFFFFFFFFFFFFFFFFFFFFFFFFFFFFFFFFFFFFFFFFFFFFFFFFFFFFFFFFFFFFFFFFFFF
%FIGURE 7
%FFFFFFFFFFFFFFFFFFFFFFFFFFFFFFFFFFFFFFFFFFFFFFFFFFFFFFFFFFFFFFFFFFFFFFFFFFFFFFFF
\begin{figure*}
\begin{center}
\includegraphics[width=75mm,angle=0]{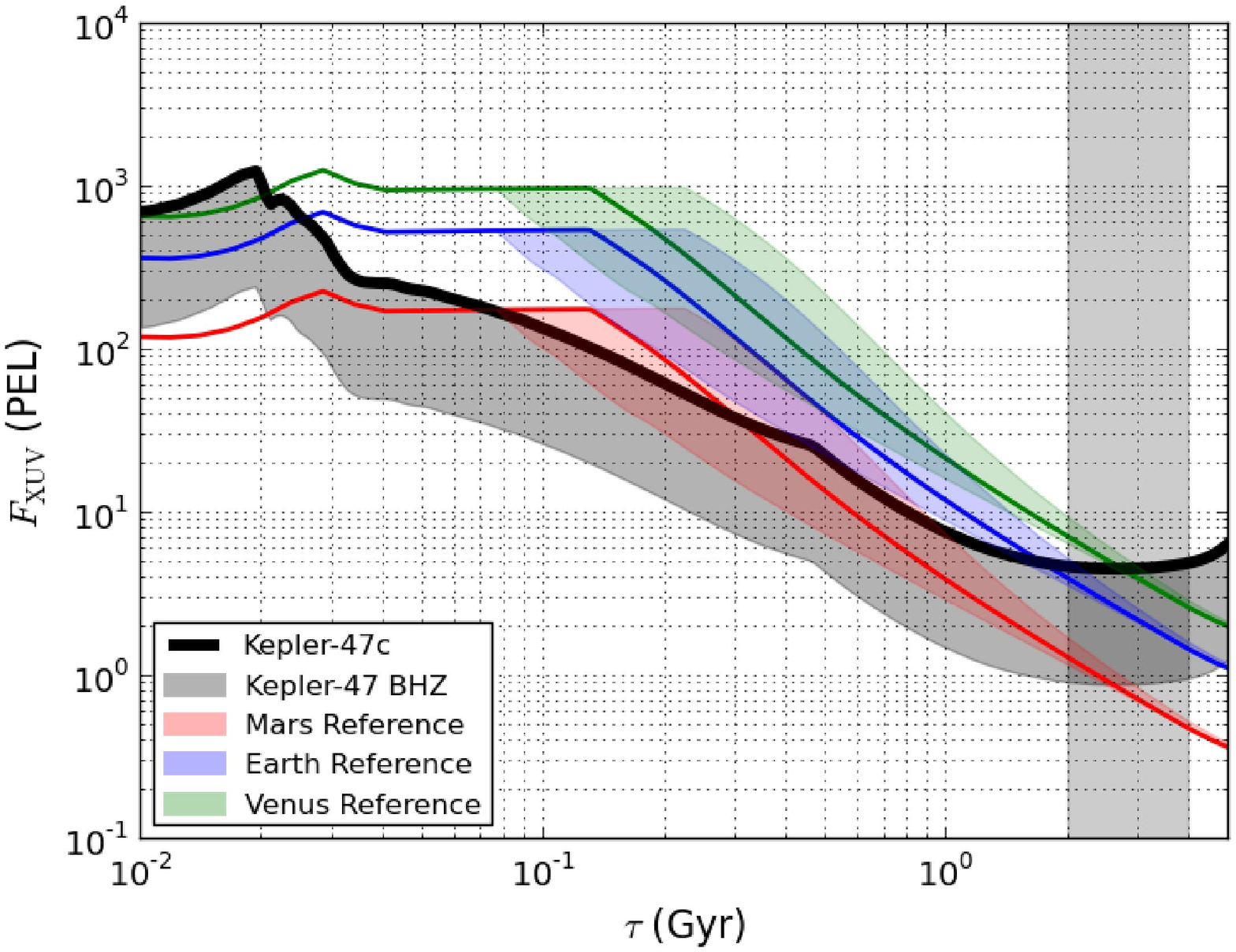}\hspace{0.5cm}%DONE
\includegraphics[width=75mm,angle=0]{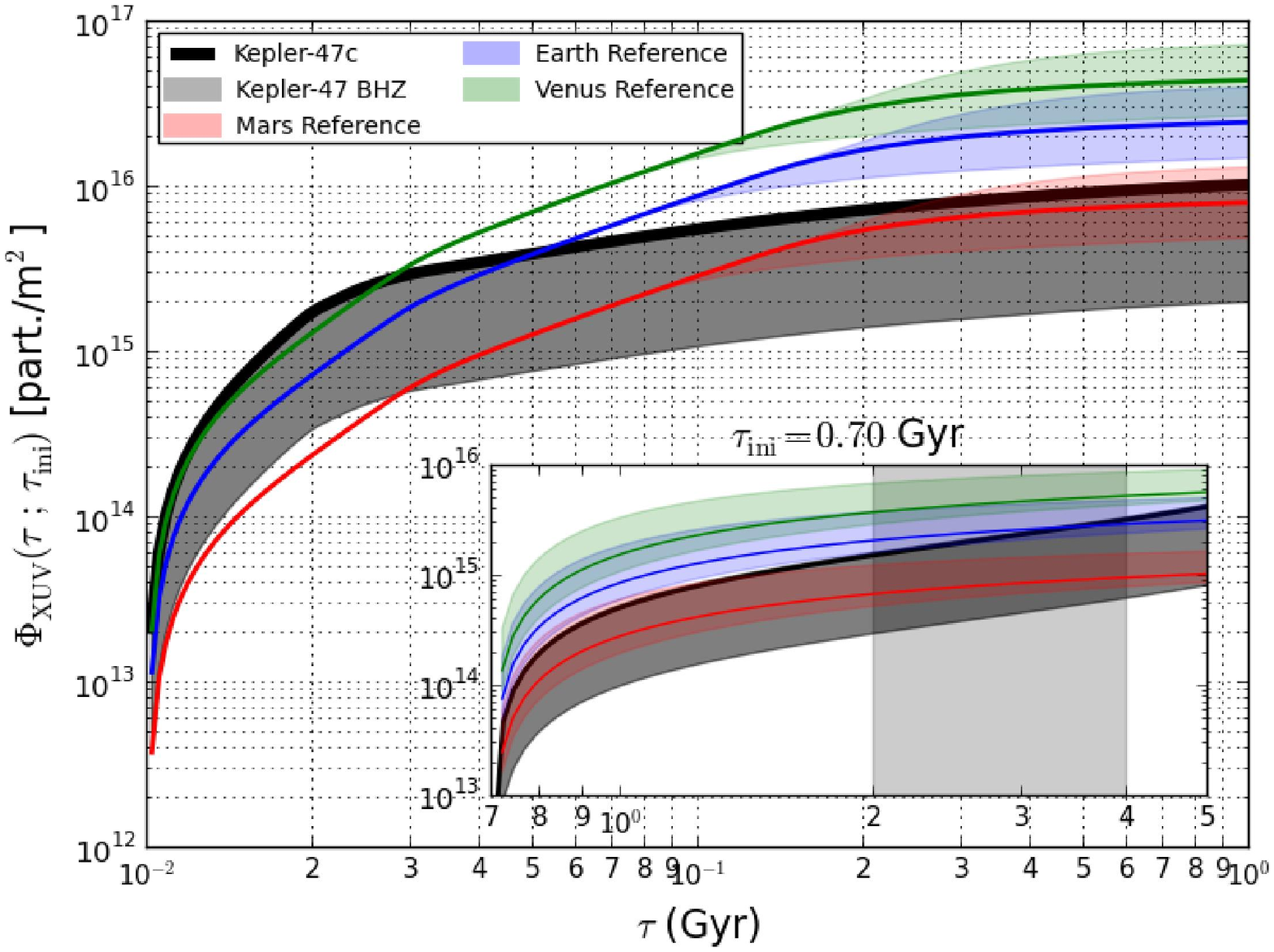}\\\vspace{0.2cm}%DONE
\includegraphics[width=75mm,angle=0]{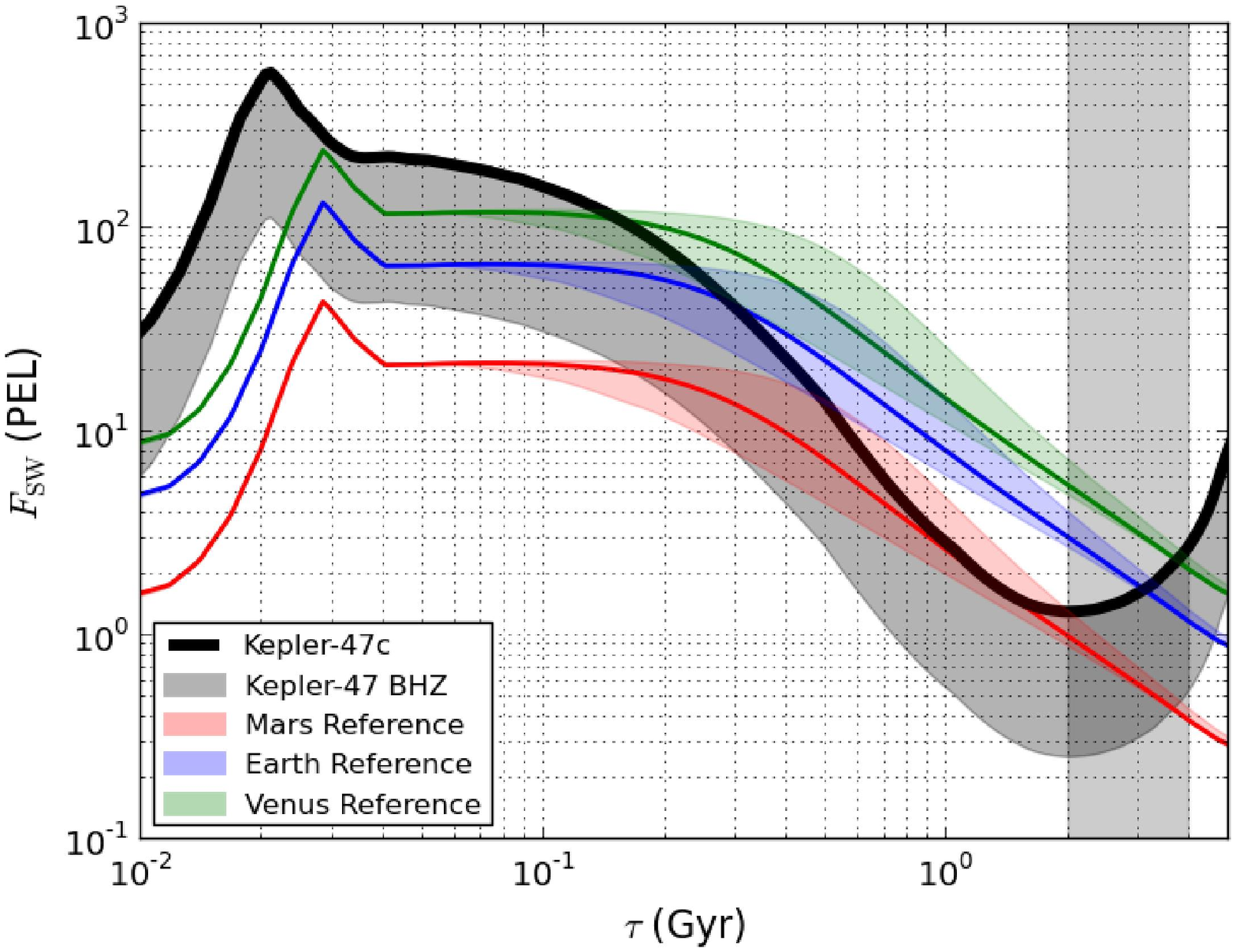}\hspace{0.5cm}%DONE
\includegraphics[width=75mm,angle=0]{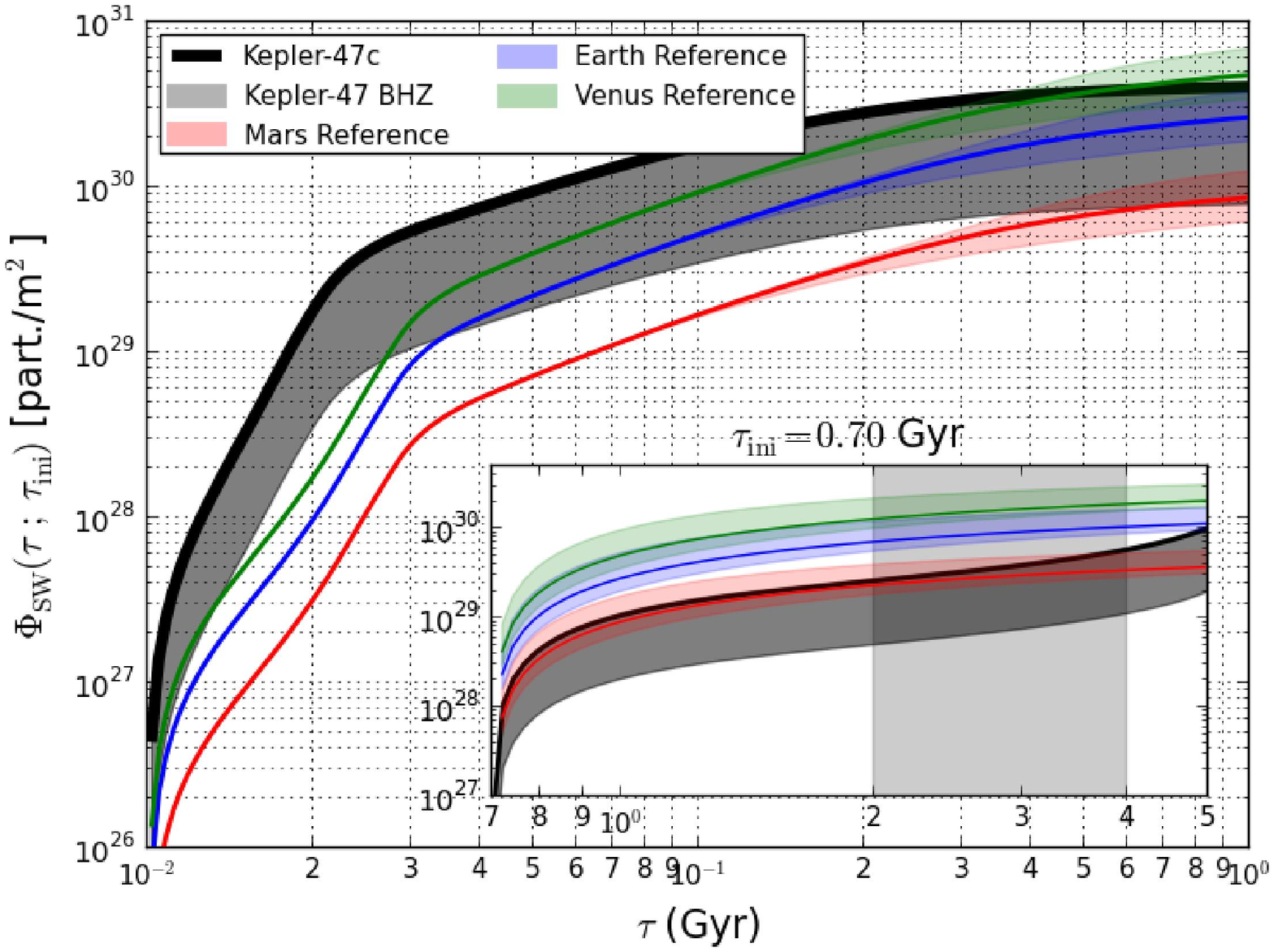}%DONE
\caption{SW and XUV flux evolution for Kepler-47.  Reduced XUV flux
  conditions early (top left panel) result in low levels of integrated
  XUV flux (top right panel). However, high SWs early (bottom left
  panel) result in a Venus-like integrated wind calculation at late
  times (bottom right panel).
\label{fig:SWXUV-Kepler47}}
\end{center}
\end{figure*}
%FFFFFFFFFFFFFFFFFFFFFFFFFFFFFFFFFFFFFFFFFFFFFFFFFFFFFFFFFFFFFFFFFFFFFFFFFFFFFFFF

The case of Kepler-47 is also very interesting (see Figure
\ref{fig:SWXUV-Kepler47}).  In this case, both radiation and plasma
environments, at least with respect to the evolution of secondary
atmospheres (inset panels), are more planet friendly than \hl{most of
  the solar system HZ}. The reason, however, is different in Kepler-47
than in Kepler-16.

First, the primary star is slightly more massive than the Sun and
produces more XUV radition. It loses mass at almost double the solar
rate during the very early phases of stellar evolution \hl{(lower left
  panel in Figure \ref{fig:SWXUV-Kepler47})}.  The saturation phase,
\hl{i.e. the phase during which Rossby number is low and chromospheric
  activity saturates (see plateau in XUV flux in the middle panel of
  Figure \ref{fig:SolarReference})}, ends at an early time (20 Myr as
compared to 100 Myr in the case of the Sun).  As a result, the XUV
luminosity decreases much more rapidly during the first couple of
hundreds of Myr.  This provides an advantage to any protoplanetary and
even secondary atmosphere \hl{developed during that phase of stellar
  and planetary evolution (which extends almost until the age of 1
  Gyr).  In the long run, any planet in the HZ of Kepler-47 will have
  experienced, by the age of the solar system, a cumulative XUV flux
  lower than most anywhere within the Sun's HZ.}

Around 20 Myr, tidal interaction starts to significantly affect
primary rotation \hl{(see middle panel of Figure
  \ref{fig:RotationEvolutionKBHZ}).  The effects do not accumulate,
  however, until the stars reach $\sim$1 Gyr}.

It is interesting to notice that Kepler-47 offers only enhanced
conditions in terms of XUV flux early.  If our assumptions are
correct, at around 50 Myr, an Earth-like planet in the inner edge of
the BHZ will have accumulated XUV flux at a level lower than Venus. In
the SW case, integrated fluxes during the first several hundreds of
Myr are enhanced with respect to the Sun.  \hl{However, if secondary
  atmospheres are common, Mars-sized objects placed at any stable
  position within the BHZ could preserve their atmospheres.}

%%%%%%%%%%%%%%%%%%%%%%%%%%%%%%%%%%%%%%%%%%%%%%%%%%%%%%%%%%%%%%%%%%%%%%%%%%%%%%%%
\subsection{\hls{Kepler-453}}
\label{subsec:KIC9632895}
%%%%%%%%%%%%%%%%%%%%%%%%%%%%%%%%%%%%%%%%%%%%%%%%%%%%%%%%%%%%%%%%%%%%%%%%%%%%%%%%

%FFFFFFFFFFFFFFFFFFFFFFFFFFFFFFFFFFFFFFFFFFFFFFFFFFFFFFFFFFFFFFFFFFFFFFFFFFFFFFFF
%FIGURE 8
%FFFFFFFFFFFFFFFFFFFFFFFFFFFFFFFFFFFFFFFFFFFFFFFFFFFFFFFFFFFFFFFFFFFFFFFFFFFFFFFF
\begin{figure*}
\begin{center}
\includegraphics[width=75mm,angle=0]{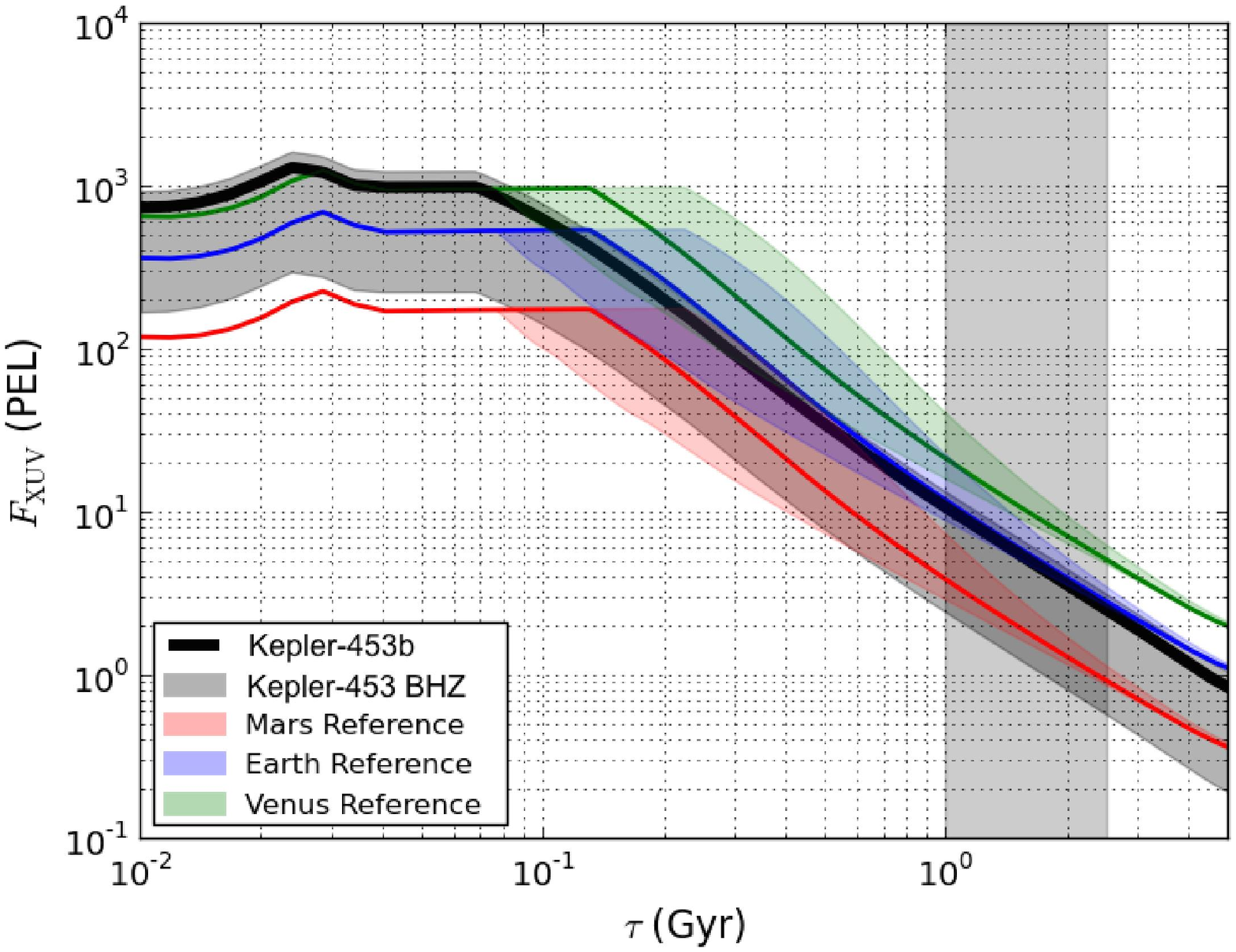}\hspace{0.5cm}%DONE
\includegraphics[width=75mm,angle=0]{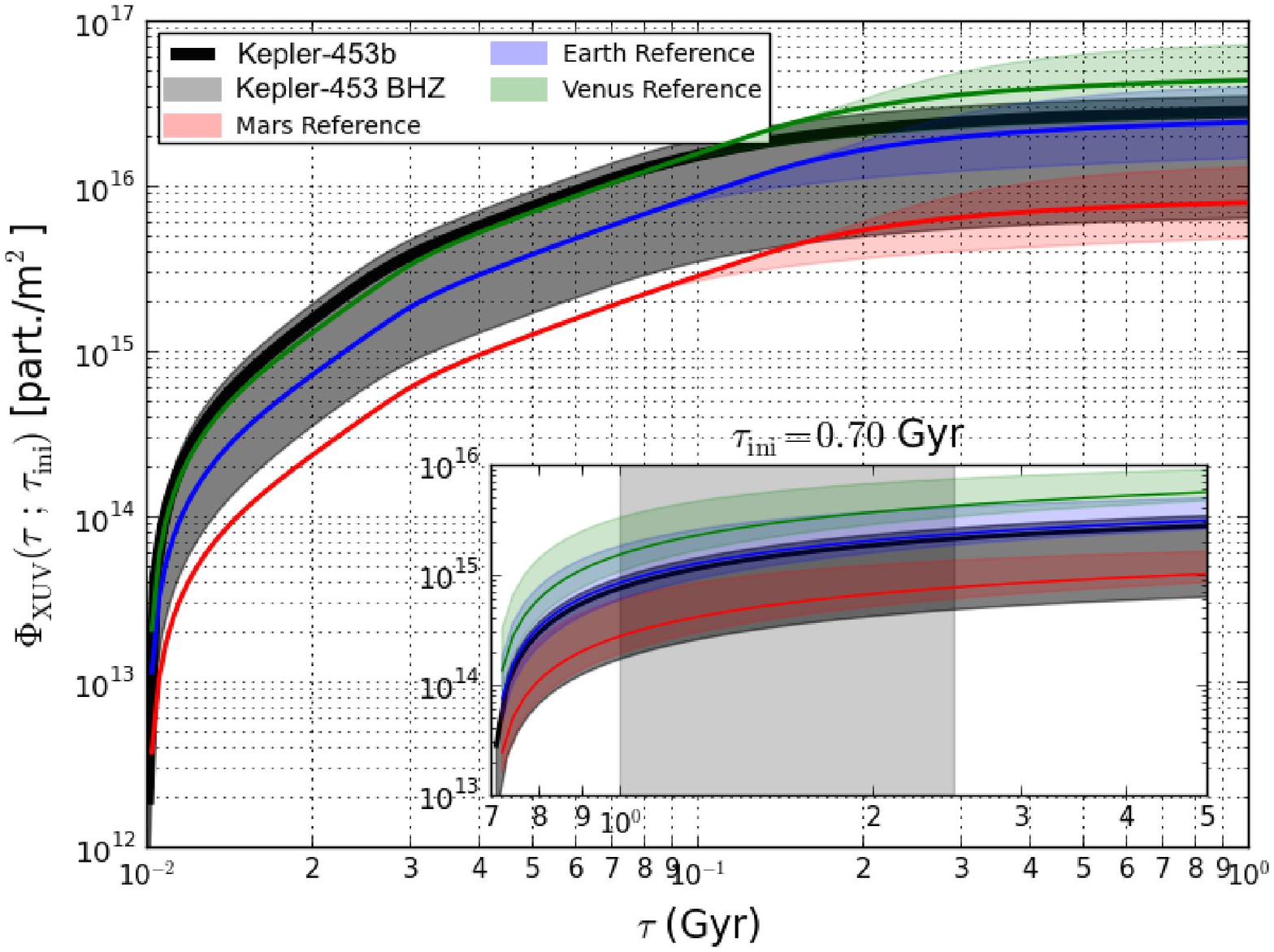}\\\vspace{0.2cm}%DONE
\includegraphics[width=75mm,angle=0]{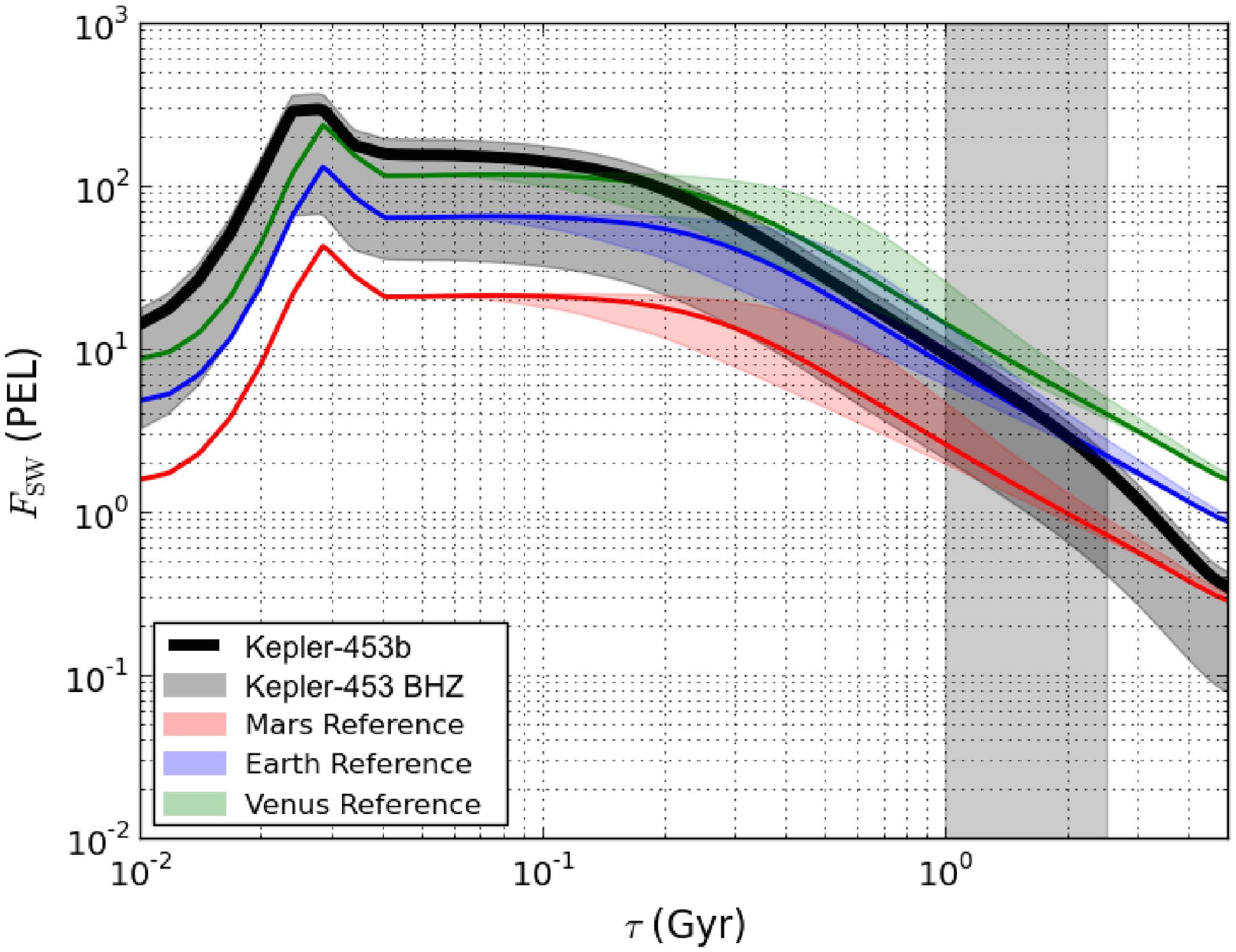}\hspace{0.5cm}%DONE
\includegraphics[width=75mm,angle=0]{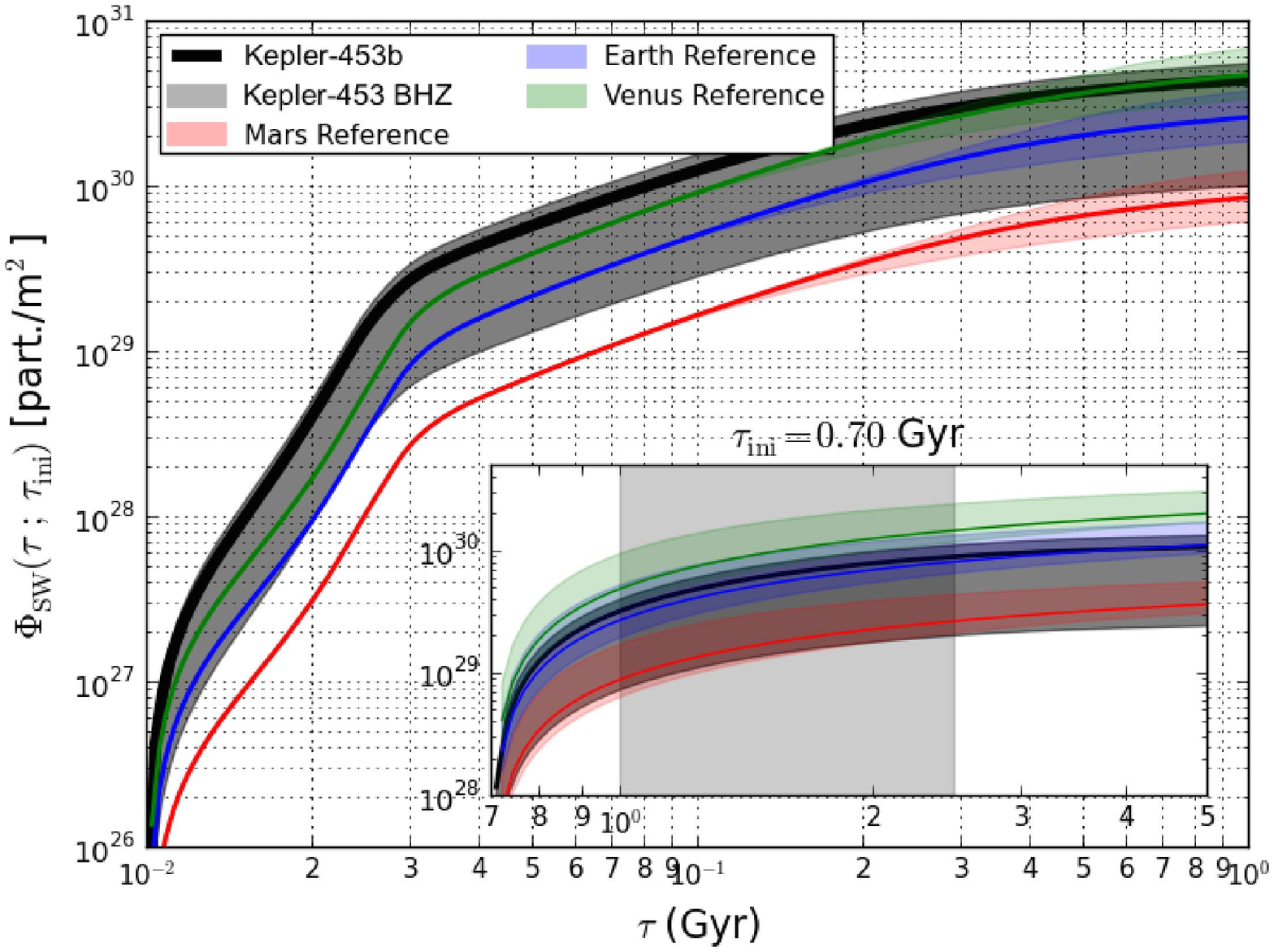}%DONE
\caption{SW and XUV flux evolution for \hls{Kepler-453}. Planets
  residing near the inner edge of the BHZ could be habitable owing to
  low integrated XUV and SW fluxes (right two panels).
  \label{fig:SWXUV-KIC9632895}}
\end{center}
\end{figure*}
%FFFFFFFFFFFFFFFFFFFFFFFFFFFFFFFFFFFFFFFFFFFFFFFFFFFFFFFFFFFFFFFFFFFFFFFFFFFFFFFF

The case of \hls{Kepler-453} (Figure \ref{fig:SWXUV-KIC9632895}) is
close to that of the solar system.  In terms of insolation, XUV, and
SW fluxes, the circumbinary system is dominated by the primary star.
The binary period of 27 days and a binary eccentricity of 0.05 (see
Table 1) \hl{produce} minimal tidal interaction for these stars.
Differences observed in the integrated XUV and SW fluxes arise
exclusively from differences in stellar properties.

The most interesting observation concerning the environment of
\hls{Kepler-453} is that even the inner edge of the BHZ could be friendly
to Venus-like planets (low integrated XUV and SW fluxes).
If the fate of Venus habitability was tightly linked to early solar
activity, \hls{Kepler-453} could theoretically harbor more than one
Earth-like planet within its BHZ.

%%%%%%%%%%%%%%%%%%%%%%%%%%%%%%%%%%%%%%%%%%%%%%%%%%%%%%%%%%%%%%%%%%%%%%%%%%%%%%%%
\section{Summary and conclusions}
\label{sec:SummaryConclusions}
%%%%%%%%%%%%%%%%%%%%%%%%%%%%%%%%%%%%%%%%%%%%%%%%%%%%%%%%%%%%%%%%%%%%%%%%%%%%%%%%

We have developed a comprehensive and largely consistent model to
calculate the evolution of stellar activity in binaries, aimed at
constraining the radiation and plasma environment of BHZ planets. We
apply the model to the {\it Kepler} binaries with known planets in the
BHZ, the {\it KBHZ sample}, namely Kepler-16b, Kepler-47c and
Kepler-453b.

We find that Kepler-16 probably has a hospitable plasma environment
for the retention of atmospheres of Mars-sized planets and exomoons in
the BHZ.  Our model predicts that the integrated SW flux anywhere in
the BHZ of the system is lower than that measured at the distance of
Mars in the solar system.  The levels of high-energy radiation are,
however, similar to those observed in the solar system.

Kepler-47 is the only system among the three in the KBHZ sample where
tidal interaction has shaped the recent history of stellar activity.
High-energy (XUV) radiation levels are affected the most by the
evolution of rotation.  The historical levels of XUV radiation
predicted by our model for this system are lower than those measured
in the solar system.  This result hints at an intriguing possibility
that Earth-like atmospheres could survive desiccation at distances
analogous to that of Venus in the solar system.  The predicted plasma
flux in this system is, however, too high for Mars-sized planets to
retain their primary atmospheres.  If, however, other mechanisms are
responsible for the degassing of a secondary atmosphere, the late
integrated fluxes seem to be lower than those predicted in the solar
system.

Finally, the case of Kepler-453 seems to be rather conventional.  The
circumbinary plasma and radiation environment is quite similar to that
found in the solar system.  However, at late times, the integrated
high-energy radiation as measured at the inner edge of the HZ (where
the actual planet Kepler-453b lies) could be better than that
experienced by Venus.

\hl{These results are not definitive, but they provide a solid
  starting point to consistently assess the key problem of the
  radiation and plasma environment of planets in binaries.  Rotational
  evolution models of stars in binaries need to be further improved
  and, more importantly, compared with a larger sample of observed
  rotations in binaries.  Models of the relationship between stellar
  activity and fundamental properties and rotation period of stars,
  across a wide range of stellar masses, need to be further developed
  and tested.  We cannot rely only on age-activity relationships found
  for single-stars in order to assess the evolution of activity of
  stars in binaries.  Stellar rotation is modified by mutual tidal
  interaction. Any future improvement of these aspects of the models
  can be tested using the computational tool we have developed and
  made avaiable at \url{http://bhmcalc.net}.}

\section*{Acknowledgments}

We thank W. Welsh for discussions concerning the {\it Kepler} planets
and N. Haghighapor for pointing out the importance of resonances.  We
also thank S. Cranmer, G. Torres, and A. Claret for discussion about
their models applied in this work.  We appreciate the help of
I. Baraffe who provided detailed stellar evolutionary models,
including moment-of -nertia calculations, required in the rotational
evolution calculations.  \hl{Special thanks to the anonymous referee,
  who provided useful input for improving the BHZ models and
  suggestions for improving the clarity of the manuscript}.
J.I. Zuluaga and P.A. Cuartas are supported by Estrategia de
Sostenibilidad 2014-2015 de la Universidad de Antioquia and by
COID/UdeA. P.A.M. is supported by NMSU-DACC.  J.I. Zuluaga is also
supported by the Fulbright Commission, Colombia.  He also thanks the
Harvard-Smithsonian Center for Astrophysics for their hospitality.

\end{document}